\documentclass[a4paper,oneside,10pt]{article}%
\usepackage{amsmath}
\usepackage{amsfonts}
\usepackage{amssymb}
\usepackage{graphicx}
\usepackage[square,numbers,sort&compress]{natbib}
\usepackage{color}%
\providecommand{\U}[1]{\protect\rule{.1in}{.1in}}

\newtheorem{theorem}{Theorem}[section]

\newtheorem{corollary}[theorem]{Corollary}

\newtheorem{definition}[theorem]{Definition}
\newtheorem{assumption}[theorem]{Assumption}
\newtheorem{example}[theorem]{Example}

\newtheorem{proposition}[theorem]{Proposition}
\newtheorem{remark}[theorem]{Remark}

\numberwithin{equation}{section}

\newcommand{\E}{{\mathbb E}}
\newcommand{\R}{{\mathbb R}}
\newcommand{\pf}{\noindent\textbf{Proof:} }
\newcommand{\eof}{\hfill{$\Box$}}

\newcommand{\argmin}{\ensuremath{\operatorname*{argmin}}}
\newcommand{\BMO}{\mathcal{BMO}}
\begin{document}

\title{Mean-variance portfolio selection with non-linear wealth
dynamics and random coefficients}
\author{Shaolin Ji \thanks{Zhongtai Securities Institute for Financial Studies,
Shandong University, Jinan 250100, China; Email: jsl@sdu.edu.cn. This work is
supported by National Natural Science Foundation of China (No. 11971263);
Supported by the Programme of Introducing Talents of Discipline to
Universities of China (No.B12023). }
\and Hanqing Jin\thanks{Mathematical Institute and Oxford-Man Institute of
Quantitative Finance, The University of Oxford, Woodstock Road, Oxford OX2
6GG, UK; Email: jinh@maths.ox.ac.uk}
\and Xiaomin Shi\thanks{Corresponding author. School of Statistics and  Mathematics, Shandong University of Finance and Economics, Jinan
250100, China; Email: shixm@mail.sdu.edu.cn. This work is supported by National Natural Science Foundation of China (No. 11801315); Supported by Natural
Science Foundation of Shandong Province (No. ZR2018QA001)}}
\maketitle

\textbf{Abstract}. This paper studies the continuous time mean-variance
portfolio selection problem with one kind of non-linear wealth dynamics. To deal the expectation constraint,  an auxiliary stochastic control problem is firstly solved by two new generalized stochastic Riccati equations from which a candidate portfolio in feedback form is constructed, and the corresponding wealth process will never cross the vertex of the parabola. In order to verify the optimality of the candidate portfolio, the convex duality  (requires the monotonicity of the cost function) is established to give another more direct expression of the terminal wealth level.  The variance-optimal martingale measure and the link between the non-linear financial market and the classical linear market are also provided. Finally, we obtain the efficient frontier  in closed form. From our results, people are more likely to invest their money in riskless asset compared with the classical linear market.

{\textbf{Key words}. } mean-variance portfolio selection; non-linear wealth
dynamic; Riccati equation; convex duality; variance-optimal martingale measure

\textbf{Mathematics Subject Classification (2010)} 60H10 93E20

\addcontentsline{toc}{section}{\hspace*{1.8em}Abstract}

\section{Introduction}
A mean-variance portfolio selection problem is to find the optimal portfolio
strategy which minimizes the variance of its terminal wealth while its
expected terminal wealth equals a prescribed level. Markowitz \cite{Ma1},
\cite{Ma2} first studied this problem in the single-period setting. Its
multi-period and continuous time counterparts have been studied extensively in
the literature; see, e.g. Bielecki et al. \cite{BJPZ}, Jin et al. \cite{JYZ}, Li et al. \cite{LN}, Li et al. \cite{LZL},
Zhou et al. \cite{ZL} and the references therein.
For the general topic of mean variance hedging, please refer to $\mathrm{\breve{C}}$ern$\mathrm{\acute{y}}$ et al. \cite{CCK}, $\mathrm{\breve{C}}$ern$\mathrm{\acute{y}}$  and Kallsen  \cite{CKa}, Schweizer \cite{Sc2010}.

Most of the literature on mean-variance
portfolio selection stays in a linear market, i.e., the wealth dynamic is a
linear equation due to the proper market setting like frictionless trading.
While in reality, the wealth dynamic is rare to be linear because of
different kinds of friction in trading, and we have to deal with nonlinearity
in the market. For example, a large investor's portfolio  may affect
the return of the stock's price which leads to a non-linear wealth dynamic.
When some taxes must be paid on the gains made on the stocks, we also
encounter nonlinearity in the wealth equation.

As for the continuous time mean-variance portfolio selection problem with
non-linear wealth dynamic, Ji \cite{Ji} obtained a necessary condition for the
optimal terminal wealth when the drift of the wealth dynamic is differentiable.
 He derived a stochastic maximum
principle which characterized the optimal terminal wealth. But the stochastic maximum
principle in Ji \cite{Ji} relies heavily on the differentiability  assumption
of the drift with respect to $(X,\pi)$. For our non-differentiable case, the key step is to find an
appropriate sub-derivative so as to construct the optimal wealth which is not concerned in \cite{Ji}.
Fu et al. \cite{FLL} studied the continuous time mean-variance portfolio selection
problem with higher borrowing rate in which the wealth dynamic is non-linear
and the coefficient is not smooth. They employed the viscosity solution of the
HJB equation to characterize the optimal portfolio strategy.

In this paper, the continuous time mean-variance portfolio selection problem
with one kind of non-linear wealth dynamics is studied. The drift is not differentiable with respect to $\pi$ in the model. When the
coefficients are all deterministic continuous functions, Ji and Shi \cite{JS}
solved this problem via the viscosity solution of the corresponding HJB
equation. But for non-linear wealth dynamics with random coefficients such as stochastic return rates
and stochastic volatilities, the method of HJB equation is no longer applicable.

Compared with classical linear market, the non-linear wealth dynamic brings new challenges. As the terminal expectation constraint $\E X_T^{\pi}=K$ is no longer linear in $\pi$. Whence it is unclear whether the feasible portfolio set is convex or not. Furthermore,  the Lagrange strong duality which was widely used in solving mean-variance portfolio selection problem for linear market (see e.g. \cite{HZ}, \cite{LZL}) is absent  a priori. Instead, by introducing a Lagrange multiplier,  we only have weak duality. Fortunately, we can take advantage of the weak duality to fix a lower bound for our problem, then construct a candidate portfolio $\pi$, and verify the optimality of $\pi$ finally. In this procedure, we will in the first place confront a stochastic control problem without state constrain (but with non-linear dynamic and quadratic cost). Inspired by Hu and Zhou \cite{HZ} in which the  mean-variance portfolio selection problem with  cone constraints was
studied, this stochastic control problem could be solved by
a generalized linear quadratic (LQ) approach. We find that our problem
can be solved by studying the positive and negative parts of the process
$X_{t}-de^{\int_{t}^{T}r_{s}ds}$ separately (see Theorem \ref{feedback}). This approach leads to two new generalized stochastic Riccati equations. Through  an exponential transformation,  we prove the global solvability of these two
generalized stochastic Riccati equations. Furthermore, we show that the
positive or negative of the process $X_{t}-de^{\int_{t}^{T}r_{s}ds}$ depends only on the positive or negative of its  initial value (see Remark \ref{vertex}). Things become apparently different when there are jumps in the price processes, please see Czichowsky and Schweizer \cite{CS}, where a coupled system of backward stochastic differential equations (BSDEs) is deduced to characterize the value process. Then by solving a convex optimization problem \eqref{maxmin}, a candidate  portfolio in feedback form is obtained.

But when it comes to verify the optimality of the candidate $\pi$ (mainly $\E X_T^{\pi}=K$), this feedback form is no longer friendly. So  the
convex duality method,  a theory which was highly developed in  utility maximization problems (see e.g. Cvitanic and Karatzas \cite{CK} and the seminal book \cite{KS} for a systematic account on this subject)  is applied to give another expression of the candidate portfolio and, especially, its corresponding terminal wealth. The main advantage of this method at this stage is that it can  directly identify the optimal terminal wealth by studying the corresponding dual problem.  Even though the quadratic function, that one is trying to minimise, lacks monotonicity or Inada condition used in establishing convex duality of utility maximization problem, problem \eqref{step1} (with $\hat d$ in place of $d$) is still rather close to utility maximization because the optimal wealth process $X_t$ never crosses the vertex of the parabola as suggested by Remark \ref{vertex} ex post.  Note that this is no longer the case for processes with jumps as in Czichowsky and Schweizer \cite{CS}.

Except for  expressing the optimal terminal wealth  more directly by establishing the convex duality, we obtain some new sharp results which was not discovered in the generalized LQ approach. Further, this procedure helps us to understand the non-linear wealth dynamic better. In more detail, we succeed in obtaining the variance-optimal martingale measure, a concept introduced firstly in Schweizer \cite{Sc}, from which we find the links between the non-linear financial market and classical linear market. Actually, these two kind of markets are linked by the equivalent martingale measures, also called risk-neutral measures (see \cite{HK, HP1, HP2}). It is worth to point out that the financial market in our setting is incomplete which yields infinitely many equivalent martingale measures. Based on the explicit characterization of the variance optimal martingale measure, we show that our non-linear wealth dynamic is equivalent to a linear wealth dynamic with a appropriately chosen mean excess return rate from the viewpoint of optimization. And this mean excess return rate is exactly the sub-derivative claimed in Corollary 4.4 of Ji \cite{Ji}.

This paper is organized as follows. In section 2, we formulate the problem and sketch  the idea  to solve it. Section 3 concerns the feasibility.
The generalized LQ approach is employed to solve an auxiliary stochastic control problem without state constraint  in section
4. A real valued Lagrange multiplier is found in  sections 5. In section 6, we construct and verify the optimality of a candidate portfolio. Finally,  the efficient strategy and efficient
frontier are obtained in closed forms. Some concluding remarks are given in Section 7.

\section{Formulation of the problem}

Let $W=(W^{1},...,W^{n})^{\prime}$ be a standard $n$-dimensional Brownian
motion defined on a filtered complete probability space $(\Omega
,\mathcal{F},\{\mathcal{F}_{t}\}_{t\geq0},\mathbb{P})$, where $\{\mathcal{F}%
_{t}\}_{t\geq0}$ denotes the natural filtration associated with the
$n$-dimensional Brownian motion $W$ and augmented.

We introduce the following spaces:
\[%
\begin{array}
[c]{l}%
L^{2}(\Omega,\mathcal{F}_{T};\mathbb{R})=\Big\{\xi:\Omega\rightarrow
\mathbb{R}\big|\xi\mbox { is }\mathcal{F}_{T}\mbox{-measurable, and }\E|\xi|^{2}%
<\infty\Big\},\\
L^{2}(0,T;\mathbb{R})=\Big\{\phi:[0,T]\times\Omega\rightarrow
\mathbb{R}\big|(\phi_{t})_{0\leq t\leq T}\mbox{ is an } \ \{\mathcal{F}_t\}_{t\geq0} \ \mbox{-predictable process,}\\
\mbox{  \ \ \ \  and }\E\int_{0}^{T}|\phi_{t}|^{2}dt<\infty
\Big\},\\
L^{\infty}(0, T;\mathbb{R})=\Big\{\phi:[0, T]\times\Omega
\rightarrow\mathbb{R}\;\Big|\;(\phi_{t})_{0\leq t\leq T}\mbox{ is an }\{\mathcal{F}%
_{t}\}_{t\geq0}\mbox{-predictable essentially  bounded process} \Big\}.
\end{array}
\]
These definitions are generalized in the obvious way to the cases when $\phi$ is $\mathbb{R}^m$, $\mathbb{R}^n$ or $\mathbb{R}^{m\times n}$-valued. In our argument, ``almost surely" (a.s.), ``almost everywhere" (a.e.) and
$(t,\omega)$  may be suppressed for notation simplicity in some circumstances when no confusion occurs. Throughout this paper, we take the following notations. For any $x\in
\mathbb{R}^{m}$, denote
\[
x^{+}=(x_{1}^{+},...,x_{m}^{+})^{\prime},\ x^{-}=(x_{1}^{-},...,x_{m}%
^{-})^{\prime},
\]
where
\[
x_{i}^{+}=\left\{
\begin{array}
[c]{c}%
x_{i},\text{ if }x_{i}\geq0;\\
0,\ \text{ if }x_{i}<0,
\end{array}
\right.  \ \text{and }x_{i}^{-}=(-x_{i})^{+},\ i=1,...,m.
\]
For any $\underline x, \ \bar x\in\mathbb{R}^m$, we write $\underline x\leq \bar x$ if $\underline x_i\leq\bar x_i, \ i=1,...,m$.

Consider a financial market consisting of a riskless asset (the money market
instrument or bond) whose price is $S^{0}$ and $m$ $(m\leq n)$ risky securities (the
stocks) whose prices are $S^{1},...,S^{m}$. An investor decides at time
$t\in\lbrack0,T]$ what amount $\pi_{t}^{i}$ of his total wealth $X_{t}$ to
invest in the $i$th stock, $i=1,...,m$. The portfolio $\pi_{t}=(\pi_{t}%
^{1},...,\pi_{t}^{m})^{\prime}$ and $\pi_{t}^{0}:=X_{t}-\sum_{i=1}^{m}\pi
_{t}^{i}$ are $\mathcal{F}_{t}$-adapted. Then
consider the following non-linear wealth dynamic:
\begin{equation}%
\begin{cases}
dX_{t}=(r_{t}X_{t}+(\pi_{t}^{+})^{\prime}\underline{\mu}_{t}%
-(\pi_{t}^{-})^{\prime}\bar{\mu}_{t})dt+\pi_{t}^{\prime}%
\sigma_{t}dW_{t},\\
X_{0}=x\in\R,\;t\in\lbrack0,T]
\end{cases}
\label{wealth}%
\end{equation}
where $r_{t}$ is the interest rate, $\underline{\mu}_{t}=(\underline{\mu
}_{t}^{1},...,\underline{\mu}_{t}^{m})^{\prime}$, $\bar{\mu}_{t}=(\bar{\mu}_{t}^{1},...,\bar{\mu}_{t}^{m}%
)^{\prime}$ are  mean excess return rates for long positions
and  short positions, and $\sigma_{t}=\{\sigma_{t}^{ij}\}_{1\leq
i\leq m, 1\leq j\leq n}$ is the volatility rate of risky assets.
Note that the drift of the wealth equation \eqref{wealth} is Lipschitz but not differentiable with respect to $\pi$, which violates assumption (H1) in \cite{Ji}.

\begin{assumption}\label{assu1}
$r$ is a deterministic measurable bounded scalar-valued
function.
\end{assumption}

\begin{assumption}\label{assu3}
$\underline\mu, \ \bar\mu\in L^\infty(0,T;\mathbb{R}^m)$ and $\underline{\mu}_{t}\leq\bar{\mu
}_{t},\ i=1,...,m.$ $\sigma\in L^\infty(0,T;\mathbb{R}^{m\times n})$ and
\[
\exists\varepsilon>0,\ \rho^{\prime}\sigma_t\sigma^{\prime}_t\rho
\geq\varepsilon|\rho|^{2},\ \forall\rho\in\mathbb{R}^{m}.
\]
\end{assumption}

Indeed, it is the following three examples that motivate us to study the wealth
dynamic (\ref{wealth}).   For simplicity, we suppose that there is only one
stock in each of these three examples.

\begin{example}
[Short selling is costly]\label{exm-1}
Jouini and Kallal \cite{JK,
JK01} proposed the following model.

Let  $\bar
b_{t}\geq\underline{b}_{t}\geq r_{t}$. When short selling is possible but costly, one has
different expected returns for long and short position of the stock. In this
case, the asset prices are given by
\[%
\begin{cases}
dS_{t}^{0}=S_{t}^{0}r_{t}dt,\ S_{0}^{0}=s_{0};\\
dS_{t}^{1}=S_{t}^{1}\Big[\big(\underline{b}_{t}I_{\{\pi_{t}\geq0\}}+\bar
{b}_{t}I_{\{\pi_{t}<0\}}\big)dt+\sigma_{t}dW_{t}\Big],\ S_{0}^{1}=s_{1}>0.
\end{cases}
\]

Then the wealth process $X\equiv X_{{}}^{x,\pi}$ of the self-financed investor
who is endowed with initial wealth $x$ is governed by the following
stochastic differential equation,
\[%
\begin{cases}
dX_{t}=\pi_{t}\frac{dS_{t}^{1}}{S_{t}^{1}}+(X_{t}-\pi_{t})\frac{dS_{t}^{0}%
}{S_{t}^{0}}\\
\ \ \ \ \ \ =(r_{t}X_{t}+\pi_{t}^{+}\underline{\mu}_{t}-\pi
_{t}^{-}\bar{\mu}_{t})dt+\pi_{t}\sigma_{t}dW_{t},\\
X_{0}=x,
\end{cases}
\]
where $\underline{\mu}_{t}=\underline{b}_{t}-r_{t},\ \bar{\mu}_{t}=\bar{b}_{t}-r_{t}, \ t\in[0,T]$.
\end{example}

\begin{example}
[Price pressure model for large investors]\label{exm-2}Cuoco and Cvitanic
\cite{CC} gave the following price pressure model.

Let  $\varepsilon$ be a small positive
number such that $b_{t}-r_{t}\geq\varepsilon\geq0$. The portfolio strategy of
a large investor could affect the expected return of the stock and the
affection level is small. The asset prices are governed by
\[%
\begin{cases}
dS_{t}^{0}=S_{t}^{0}r_{t}dt,\ S_{0}^{0}=s_{0};\\
dS_{t}^{1}=S_{t}^{1}\Big[\big(b_{t}-\varepsilon\, \mathrm{sgn}(\pi
_{t})\big)dt+\sigma_{t}dW_{t}\Big],\ S_{0}^{1}=s_{1}>0,
\end{cases}
\]
where
\begin{align*}
\mathrm{sgn}(x)=%
\begin{cases}
\frac{|x|}{x},\ \ \text{if}\ \ x\neq0;\\
0,\ \ \ \ \text{otherwise}.
\end{cases}
\end{align*}
In this specific large investor model, buying the risky security depresses
its expected return while shorting it increases its expected return as
explained in Cuoco and Cvitanic \cite{CC}.

The wealth equation can be written
\begin{align*}%
\begin{cases}
dX_{t} =(r_{t}X_{t}+(b_{t}-r_{t})\pi_{t}-\varepsilon|\pi_{t}|)dt+\pi_{t}%
\sigma_{t}dW_{t}\\
\ \ \ \ \ \ =(r_{t}X_{t}+\pi^{+}_{t}\underline{\mu}_{t}-\pi
_{t}^{-}\bar{\mu}_{t})dt+\pi_{t}\sigma_{t}dW_{t},\\
X_{0}=x,
\end{cases}
\end{align*}
where $\underline{\mu}_{t}:=b_{t}-r_{t}-\varepsilon$ and
$\bar\mu_{t}:=b_{t}-r_{t}+\varepsilon, \ t\in[0,T]$.
\end{example}

\begin{example}
[Trading with taxes]\label{exm-3}El Karoui et al. \cite{EPQ1} studied the
following financial model with taxes.

Let  $\alpha\in\lbrack0,1)$ be a
constant. And $b_{t}\geq r_{t}$. The asset prices are given
by
\[%
\begin{cases}
dS_{t}^{0}=S_{t}^{0}r_{t}dt,\ S_{0}^{0}=s_{0};\\
dS_{t}^{1}=S_{t}^{1}(b_{t}dt+\sigma_{t}dW_{t}),\ S_{0}^{1}=s_{1}>0.
\end{cases}
\]
And there are some taxes which must be paid on the gains made on the stock. In
this case, the wealth equation satisfies
\[%
\begin{cases}
dX_{t}=(r_{t}X_{t}+(b_{t}-r_{t})\pi_{t}-\alpha\pi^{+}(b_{t}-r_{t}))dt+\pi
_{t}\sigma_{t}dW_{t}\\
\ \ \ \ \ \ =((r_{t}X_{t}+\pi_{t}^{+}\underline{\mu}_{t}-\pi
_{t}^{-}\bar{\mu}_{t})dt+\pi_{t}\sigma_{t}dW_{t},\\
X_{0}=x,
\end{cases}
\]
where $\underline{\mu}_{t}=(1-\alpha)(b_{t}-r_{t})$ and
$\bar{\mu}_{t}=b_{t}-r_{t},  \ t\in[0,T]$.
\end{example}

\begin{remark}
When $\underline{\mu}_{t}=\bar{\mu}_{t},\ t\in\lbrack0,T],\ a.s.$, the
wealth dynamic (\ref{wealth}) degenerates to the classical linear case.
\end{remark}

\begin{definition}
A portfolio $\pi$ is said to be admissible if $\sigma^{\prime}\pi\in
L^{2}(0,T;\mathbb{R}^{n})$ and $(X,\pi)$ satisfies $(\ref{wealth})$.
\end{definition}

Note that $\sigma$ is bounded, so $\sigma^{\prime}\pi\in
L^{2}(0,T;\mathbb{R}^{n})$ is equivalent to $\pi\in
L^{2}(0,T;\mathbb{R}^{m})$.
Denote by $\mathcal{A}(x)$ the set of admissible portfolio $\pi$.

Under Assumption \ref{assu1}, for a given expectation level $K\geq x_{0}e^{\int_{0}^{T}r_{s}ds}$, consider
the following continuous time mean-variance portfolio selection problem:
\begin{align}
&  \mathrm{Minimize}\ \mathrm{Var}(X_{T})={\mathbb{E}}(X_{T}-K)^{2}%
,\nonumber\\
&  s.t.%
\begin{cases}
{\mathbb{E}}X_{T}=K,\\
\pi\in\mathcal{A}(x).
\end{cases}
\label{optm}%
\end{align}

Denote $\Pi=\{\pi|\pi\in\mathcal{A}(x), \ \mbox{and} \ \E X_T=K\}$.
The problem (\ref{optm}) is called feasible if $\Pi$ is non empty.  Any $\pi\in\Pi$   is called a feasible portfolio for the problem \eqref{optm}. Denote by $X^{\pi}$ be the wealth process \eqref{wealth}  whenever it is necessary to indicate its dependence on $\pi\in\mathcal{A}(x)$. A optimal strategy $\pi^{\ast}$ to (\ref{optm})
is called an efficient strategy corresponding to $K$. Then $(\mathrm{Var}(X_{T}^{\pi^\ast}),K)$ is
called an efficient point. The set of all efficient points $\{(\mathrm{Var}%
(X_{T}^{\pi^\ast}),K)\mid K\in\lbrack xe^{\int_{0}^{T}r_{s}ds},+\infty)\}$ is
called the efficient frontier.

To deal with the constraint ${\mathbb{E}}X_{T}=K$, we introduce a Lagrange
multiplier $-2\lambda\in\mathbb{R}$ and obtain the following unconstrained
optimization problem:
\begin{align}
&  \inf_{\pi\in\mathcal{A}(x)}\Big[ {\mathbb{E}}(X_{T}-K)^{2}-2\lambda({\mathbb{E}}%
X_{T}-K)\Big].
\label{optmau}%
\end{align}
The problem \eqref{optmau} yields a lower bound on our original problem \eqref{optm}. To be more precisely, we have the following weak duality between problems \eqref{optm} and \eqref{optmau}:
\begin{align}\label{weakdual}
\sup_{\lambda\in\R}\inf_{\pi\in\mathcal{A}(x)}\Big[\E(X_T^{\pi}-K)^2-2\lambda(\E X_T^{\pi}-K)\Big]\leq \inf_{\pi\in\Pi}\E(X_T^{\pi}-K)^2.
\end{align}
In fact, let $\hat\pi\in\Pi$
be any feasible portfolio  and  $\lambda\in\R$, we have
\begin{align*}
\E(X_T^{\hat\pi}-K)^2-2\lambda(\E X_T^{\hat\pi}-K)= \E(X_T^{\hat\pi}-K)^2.
\end{align*}
Hence
\begin{align*}
\inf_{\pi\in\mathcal{A}(x)}\Big[\E(X_T^{\pi}-K)^2-2\lambda(\E X_T^{\pi}-K)\Big]&\leq \E(X_T^{\hat\pi}-K)^2-2\lambda(\E X_T^{\hat\pi}-K)\\
&=\E(X_T^{\hat\pi}-K)^2,
\end{align*}
for any $\lambda\in\R$ and any feasible portfolio $\hat\pi\in\Pi$. Then the weak duality \eqref{weakdual} follows.

Note that we only have the weak duality \eqref{weakdual} between problems (\ref{optm}) and
(\ref{optmau}). If the inequality becomes equality, we say that strong duality holds. And the problem in the left-hand side (LHS) of \eqref{weakdual} is more likely to be solved than our original problem \eqref{optm} (equivalently  the right-hand side (RHS) of \eqref{weakdual}). Actually, for any $d\in\R$, the problem \eqref{optmau} is a stochastic control problem without state constraint (even with non-linear dynamic), we can solve it by a generalization of linear quadratic control technique. And by denoting
\begin{align}\label{def:g}
\ell(\lambda)=\inf_{\pi\in\mathcal{A}(x)}\Big[\E(X_T^{\pi}-K)^2-2\lambda(\E X_T^{\pi}-K)\Big], \ \lambda\in\R,
\end{align}
then $\ell$ is a concave function as it is the infimum of a class of linear functions of $\lambda$. So it is not hard to solve the convex optimization problem  $\sup_{\lambda\in\R}\ell(\lambda)$.
But unfortunately, due to the non-linear wealth dynamic \eqref{wealth}, it is very difficult to establish the strong duality or even to prove the  convexity of the set of feasible portfolios $\Pi$. Nevertheless, we can still take advantage of the weak duality \eqref{weakdual} to construct a candidate  portfolio $\pi^*$ for our original problem \eqref{optm}, then verify the optimality of $\pi^*$.  The main idea is as follows:
\begin{itemize}
\item Step 1: For any $\lambda\in\R$, find a optimal portfolio $\pi^{\lambda}$ to the problem
    \eqref{optmau}.
\item Step 2: Find a argument maximum $\hat\lambda\in\R$    of
   \begin{align}\label{maxmin}
    \sup_{\lambda\in\R}\ell(\lambda).
    \end{align}
\item Step 3: Set $\pi^*=\pi^{\hat\lambda}$, then
\begin{align*}
\sup_{\lambda\in\R}\inf_{\pi\in\mathcal{A}(x)}\Big[\E(X_T^{\pi}-K)^2-2\lambda(\E X_T^{\pi}-K)\Big]&=\sup_{\lambda\in\R}\Big[\E(X_T^{\pi^\lambda}-K)^2-2\lambda(\E X_T^{\pi^\lambda}-K)\Big]\\
&=\E(X_T^{\pi^*}-K)^2-2\hat\lambda(\E X_T^{\pi^*}-K).
\end{align*}
At this time, if we can show $\pi^*\in\Pi$, i.e. $\pi^*\in\mathcal{A}(x)$ and $\E X_T^{\pi^*}=K$, then $\E(X_T^{\pi^*}-K)^2$ attains the lower bound of the original problem \eqref{optm}, i.e. the LHS of \eqref{weakdual}, which verifies the optimality of $\pi^*$ for problem \eqref{optm}.
\end{itemize}

%

\section{Feasibility}
Let us address ourselves to the feasibility of problem (\ref{optm}) firstly.

\begin{theorem}
Under Assumptions \ref{assu1} and \ref{assu3},
the mean-variance problem (\ref{optm}) is feasible for any $K\in\lbrack
xe^{\int_{0}^{T}r_{s}ds},+\infty)$ if and only if
\begin{equation}
\sum_{i=1}^{m}{\mathbb{E}}\left[  \int_{0}^{T}(\underline{\mu}_{t}^{i}%
)^{+}dt\right]  >0\text{ or }\sum_{i=1}^{m}{\mathbb{E}}\left[  \int_{0}%
^{T}(\bar{\mu}_{t}^{i})^{-}dt\right]  >0. \label{feasible}%
\end{equation}

\end{theorem}

\noindent\textbf{Proof:} (1) We first prove the \textquotedblleft if" part.

Define
\[
M_{i}=\{(t,\omega):\underline{\mu}_{t}^{i}>0\},\ i=1,2,...,m.
\]
If $\sum_{i=1}^{m}{\mathbb{E}}\left[  \int_{0}^{T}(\underline{\mu}_{t}%
^{i})^{+}dt\right]  >0$, then there exists an $i_{0}\in\{1,2,...,m\}$ such
that the product measure (in terms of $\mathbb{P}$ and the Lebesgue measure) of
$M_{i_{0}}$ is nonzero. Denote the $i_{0}^{th}$ row of $\sigma_{t}$ by
$\sigma_{t}^{i_{0}}=(\sigma_{t}^{i_{0},1},\cdots,\sigma_{t}^{i_{0},n})$ and
the length of the vector $\sigma_{t}^{i_{0}}$ by $|\sigma_{t}^{i_{0}}|$. Since
$\sigma_{t}$ is invertible, it is obvious that $|\sigma_{t}^{i_{0}}|>0$. Set
\[
\pi_{t}^i=%
\begin{cases}
1/|\sigma_{t}^{i_{0}}|, & \text{if}\ i=i_{0}\mbox{ and }(t,\omega)\in
M_{i_{0}};\\
0, & \text{if}\ i\neq i_{0}\mbox{ or }(t,\omega)\notin M_{i_{0}}.
\end{cases}
\]
For any nonnegative real number $\beta$, we construct a portfolio $\pi
_{\beta,t}:=\beta(\pi_{t}^1,\cdots,\pi_{t}^m)^{\prime}$. $\pi_{\beta,t}$
is admissible due to $\sigma_{t}^{\prime}\pi_{\beta,t}=\beta(\sigma
_{t}^{i_{0}})^{\prime}{\mathbf{1}}_{(t,\omega)\in M_{i_{0}}}/|\sigma
_{t}^{i_{0}}|$ and
\[
(\pi_{\beta,t}^{+})^{\prime}\underline{\mu}_{t}-(\pi_{\beta,t}%
^{-})^{\prime}\bar{\mu}_{t}=\beta\frac{\underline{\mu}_{t}^{i_{0}}%
}{|\sigma_{t}^{i_{0}}|}{\mathbf{1}}_{(t,\omega)\in M_{i_{0}}}.
\]
The wealth process corresponding to $\pi_{\beta}$ at time $T$ is
\begin{align*}
X_{T}  &  =xe^{\int_{0}^{T}r_{s}ds}+\int_{0}^{T}e^{\int_{t}^{T}r_{s}%
ds}((\pi_{\beta,t})^{+})^{\prime}\underline{\mu}_{t}-((\pi_{\beta,t}%
)^{-})^{\prime}\bar{\mu}_{t})dt+\int_{0}^{T}e^{\int_{t}^{T}r_{s}ds}%
\pi_{\beta,t}^{\prime}\sigma_{t}dW_{t}\\
&  =xe^{\int_{0}^{T}r_{s}ds}+\beta\int_{0}^{T}e^{\int_{t}^{T}r_{s}ds}%
\frac{\underline{\mu}_{t}^{i_{0}}}{|\sigma_{t}^{i_{0}}|}{\mathbf{1}}_{(t,\omega)\in M_{i_{0}}}dt+\beta\int%
_{0}^{T}e^{\int_{t}^{T}r_{s}ds}\frac{\sigma_{t}^{i_{0}}}{|\sigma_{t}^{i_{0}}%
|}{\mathbf{1}}_{(t,\omega)\in M_{i_{0}}}dW_{t}.
\end{align*}
Taking expectation on both sides, we get
\[
{\mathbb{E}}X_{T}=xe^{\int_{0}^{T}r_{s}ds}+\beta{\mathbb{E}}\left[
\int_{0}^{T}e^{\int_{t}^{T}r_{s}ds}\frac{\underline{\mu}_{t}^{i_0}%
}{|\sigma_{t}^{i_{0}}|}{\mathbf{1}}_{(t,\omega)\in M_{i_{0}}}dt\right]  .
\]
Define%
\[
k={\mathbb{E}}\left[  \int_{0}^{T}e^{\int_{t}^{T}r_{s}ds}\frac{\underline{\mu
}_{t}^{i_{0}}}{|\sigma_{t}^{i_{0}}|}{\mathbf{1}}_{(t,\omega)\in M_{i_{0}}}dt\right]  .
\]
We have $k>0$ since $|\sigma_{t}^{i_{0}}|>0$ and $\underline\mu_t^{i_0}{\mathbf{1}}_{(t,\omega)\in M_{i_{0}}}  >0$. Taking $\beta
=\frac{K-xe^{\int_{0}^{T}r_{s}ds}}{k}$, we obtain ${\mathbb{E}}X_{T}=K$
which means that the problem (\ref{optm}) is feasible. For the case of
$\sum_{i=1}^{m}{\mathbb{E}}\left[  \int_{0}^{T}(\bar{\mu}_{t}^{i}%
)^{-}dt\right]  >0$, the proof is similar.

(2) Conversely, if the problem (\ref{optm}) is feasible for any $K\geq
xe^{\int_{0}^{T}r_{s}ds}$, then for a given $K_{0}>xe^{\int_{0}%
^{T}r_{s}ds}$, there exists an admissible portfolio $\pi$ such that
\[
K_{0}={\mathbb{E}}X_{T}=xe^{\int_{0}^{T}r_{s}ds}+{\mathbb{E}}\left[
\int_{0}^{T}e^{\int_{t}^{T}r_{s}ds}((\pi_{t}^{+})^{\prime}\underline{\mu}%
_{t}-(\pi_{t}^{-})^{\prime}\bar{\mu}_{t})dt\right]
\]
which leads to
\begin{equation}
{\mathbb{E}}\left[  \int_{0}^{T}e^{\int_{t}^{T}r_{s}ds}((\pi_{t}^{+})^{\prime
}\underline{\mu}_{t}-(\pi_{t}^{-})^{\prime}\bar{\mu}_{t})dt\right]  >0.
\label{ine-feasibility-1}%
\end{equation}
If (\ref{feasible}) does not hold, then we have that $\underline{\mu}_{t}<0$
and $\bar{\mu}_{t}>0$ hold simultaneously for $t\in\lbrack0,T]$. It
yields that
\[
{\mathbb{E}}\left[  \int_{0}^{T}e^{\int_{t}^{T}r_{s}ds}((\pi_{t}^{+})^{\prime
}\underline{\mu}_{t}-(\pi_{t}^{-})^{\prime}\bar{\mu}_{t})dt\right]  \leq0
\]
which contradicts (\ref{ine-feasibility-1}). This completes the proof.
\hfill{$\Box$}

\begin{remark}
When $\underline{\mu}_{t}=\bar{\mu}_{t} $,
(\ref{feasible}) degenerates to ${\mathbb{E}}\left[  \int_{0}^{T}|\underline{\mu}_{t}|^{2}dt\right]  >0$.
\end{remark}

From now on, we will assume (\ref{feasible})
holding throughout this paper.

\section{Solution for the problem \eqref{optmau}}
For any $\lambda\in\R$, set $d=K+\lambda$, then
\begin{align}\label{lambdatod}
\E(X_T-K)^2-2\lambda(\E X_T-K)=\E(X_T-d)^2-\lambda^2=\E(X_T-d)^2-(d-K)^2.
\end{align}
Therefore at this step, it suffices to solve
\begin{equation}
\mathrm{Minimize}\ {\mathbb{E}}(X_{T}-d)^{2},\ s.t.\ \pi\in\mathcal{A}(x),
\label{step1}%
\end{equation}
for any $d\in\R$.

Define the following mappings:
\begin{align*}
&H_{1,t}^*(\pi, P,\Lambda):=P\pi^{\prime}\sigma
_{t}\sigma^{\prime}_{t}\pi+2[P((\pi^{+})^{\prime}\underline{\mu
}_{t}-(\pi^{-})^{\prime}\bar\mu_{t})+\pi^{\prime}\sigma
_{t}\Lambda],\\
&H_{2,t}^*(\pi, P,\Lambda):=P\pi^{\prime}\sigma
_{t}\sigma^{\prime}_{t}\pi-2[P((\pi^{+})^{\prime}\underline{\mu
}_{t}-(\pi^{-})^{\prime}\bar\mu_{t})+\pi^{\prime}\sigma
_{t}\Lambda], \ (t,\pi,P,\Lambda)\in[0,T]\times\mathbb{R}^m\times\mathbb{R}\times\mathbb{R}^n,
\end{align*}
and
\begin{align*}
&H_{1,t}( P,\Lambda):=\inf_{\pi\in\mathbb{R}^{m}}H_{1,t}^*(\pi, P,\Lambda),\\
&H_{2,t}( P,\Lambda):=\inf_{\pi\in\mathbb{R}^{m}}H_{2,t}^*(\pi, P,\Lambda), \ (t,P,\Lambda)\in[0,T]\times\mathbb{R}\times\mathbb{R}^n.
\end{align*}

Under Assumption \ref{assu3}, for any $P>0$, $\Lambda\in\mathbb{R}^n$, there exists $C_1(P,\Lambda)>0$
\begin{align*}
H_{1,t}^*(\pi, P,\Lambda)\geq \varepsilon P |\pi|^2-C_1(P+|\Lambda|)|\pi|=\varepsilon P|\pi|(|\pi|-\frac{C_1(P+|\Lambda|)}{\varepsilon P}).
\end{align*}
If $|\pi|>\frac{C_1(P+|\Lambda|)}{\varepsilon P}$, then $H_{1,t}^*(\pi, P,\Lambda)>0$. Notice that $\inf_{\pi\in\mathbb{R}^{m}}H_{1,t}^*(\pi, P,\Lambda)\leq H_{1,t}^*(0, P,\Lambda)=0$, this implies that
\begin{align*}
H_{1,t}( P,\Lambda)=\inf_{|\pi|\leq\frac{C_1(P+|\Lambda|)}{\varepsilon P}}H_{1,t}^*(\pi, P,\Lambda)>-\infty.
\end{align*}
Therefore $H_{1,t}( P,\Lambda)$ is finite. This same is true for $H_{2,t}( P,\Lambda)$.

In order to solve the sub-problem (\ref{step1}), we introduce the following
two stochastic Riccati equations:
\begin{equation}%
\begin{cases}
dP_{1,t}=-[2r_{t}P_{1,t}+H_{1,t}(P_{1,t},\Lambda_{1,t})]dt+\Lambda
_{1,t}^{\prime}dW_{t},\\
P_{1,T}=1,\\
P_{1,t}>0;
\end{cases}
\label{Riccati1}%
\end{equation}

\begin{equation}
\label{Riccati2}%
\begin{cases}
dP_{2,t}=-[2r_{t}P_{2,t}+H_{2,t}(P_{2,t},\Lambda_{2,t})]dt+\Lambda^{\prime
}_{2,t}dW_{t},\\
P_{2,T}=1,\\
P_{2,t}>0.
\end{cases}
\end{equation}
These are two BSDEs whose solutions happen to be in the class of martingales of bounded mean oscillation, briefly called BMO martingales. Here we recall some facts about this theory, see Kazamaki \cite{Ka}. The process $\int_0^\cdot \Lambda_s'dW_s$ is a BMO martingale if and only if there exists a constant $C>0$ such that
\[
\mathbb{E}\bigg[\int_\tau^T|\Lambda_s|^2ds\;\bigg|\;\mathcal F_\tau\bigg]\leq C
\]
for all stopping times $\tau\leq T$.
The stochastic exponential $\mathcal E(\int_0^\cdot \Lambda_s'dW_s)$ of a BMO martingale $\int_0^\cdot \Lambda_s'dW_s$ is a uniformly integrable martingale. Moreover, if $\int_0^\cdot \Lambda_s'dW_s$ and $\int_0^\cdot Z_s'dW_s$ are both BMO martingales, then under the probability measure $\widetilde{\mathbb{P}}$ defined by $\frac{d\widetilde{\mathbb{P}}}{d\mathbb{P}}=\mathcal E (\int_0^T Z_s'dW_s)$, $\widetilde W_t=W_t-\int_0^t Z_sds$ is a standard Brownian motion, and $\int_0^\cdot \Lambda_s'd\widetilde W_s$ is a BMO martingale.
Set
\[
\BMO=\{\Lambda\in L^2(0,T;\mathbb{R}^n)\big|\int_0^\cdot\Lambda_s'dW_s \ \mbox{is a BMO martingale}\}.
\]

\begin{definition}
\label{defsolution} A pair of processes $(P_{1},\Lambda_{1})\in L^{\infty
}(0,T;\mathbb{R})\times \BMO$ (resp.
$(P_{2},\Lambda_{2})$) is called a solution to the Riccati equation
(\ref{Riccati1}) (resp. (\ref{Riccati2})) if it satisfies ($\ref{Riccati1}$)
(resp. (\ref{Riccati2})).
\end{definition}

The Riccati equations (\ref{Riccati1}) and (\ref{Riccati2}) are highly
non-linear BSDEs which violate both the standard Lipschitz conditions and the
quadratic growth conditions. There are several results on the solvability of
stochastic Riccati equations (see for example Hu and Zhou \cite{HZ}, Kohlmann and Tang \cite{KT}). But up to
our knowledge, no results can be directly applied to (\ref{Riccati1}) and
(\ref{Riccati2}).

We first give the boundedness results of the solutions to (\ref{Riccati1}) and
(\ref{Riccati2}), which is useful in Corollary \eqref{cobon}.

\begin{proposition}
\label{boundedness} Under  Assumptions \ref{assu1} and \ref{assu3}, if $(P,\Lambda)$ is a solution to equation (\ref{Riccati1}%
) (or (\ref{Riccati2})), then
\[
P_t\le e^{2\int_{t}^{T}r_{s}ds}.
\]
\end{proposition}

\pf
 We only prove the claim for (\ref{Riccati1}) and the
proof for (\ref{Riccati2}) is similar.

Set
\[
\bar{P}_t=P_te^{2\int_{0}^{t}r_{s}ds} \ \text{ and } \ \bar{\Lambda}_t=\Lambda
_{t}e^{2\int_{0}^{t}r_{s}ds}.
\]
Then $(\bar{P},\bar{\Lambda})$ is a solution to the BSDE
\begin{align*}%
\begin{cases}
d\bar{P}_t=-e^{2\int_{0}^{t}r_{s}ds}H_{1,t}(e^{-2\int_{0}^{t}r_{s}ds}\bar
{P}_t,e^{-2\int_{0}^{t}r_{s}ds}\bar{\Lambda}_t)dt+\bar{\Lambda}^{\prime
}_tdW_{t},\\
\bar{P}_T=e^{2\int_{0}^{T}r_{s}ds},\\
\bar{P}_t>0.
\end{cases}
\end{align*}
Since $H_{1,t}\leq0$, $\bar{P}_t$ is a sub-martingale. Thus, $\bar{P}%
_t\leq{\mathbb{E}}[\bar{P}_T|\mathcal{F}_t]=\bar{P}_T$ which leads to $P_t\leq
e^{2\int_{t}^{T}r_{s}ds}$.
\eof

Now we prove the existence and uniqueness of solutions to (\ref{Riccati1}) and
(\ref{Riccati2}).

Hereafter, we shall use $C$ to represent a generic positive constant which can be different from line to line.

\begin{theorem}
\label{riccatipr}
Suppose $r\in L^\infty(0,T;\mathbb{R})$ and  Assumption \ref{assu3} hold,
there exists a unique solution $(P_{1},\Lambda_{1})$ (resp.
$(P_{2},\Lambda_{2})$) to ($\ref{Riccati1}$) (resp. ($\ref{Riccati2}$)), such that $P_1\geq C$  (resp. $P_2\geq C$) for some $C>0$.
\end{theorem}

\pf
We only prove the assertion  for ($\ref{Riccati2}$) and
the arguments for ($\ref{Riccati1}$) are analogous or obvious. The idea is to turn the stochastic Riccati equation \eqref{Riccati2} to a quadratic BSDE (through an exponential transformation) whose existence and uniqueness are known.

Set
\begin{align}\label{mathcalB}
\mathcal{B}=\{v:[0, T]\times\Omega
\rightarrow\mathbb{R}^m\mid & v\in L^\infty(0,T;\mathbb{R}^m) \ \mbox{and}\ \underline{\mu}_{t}\leq v_{t}\leq\bar{\mu}_{t}\}.
\end{align}
Recall the definition of $H_{2}(P,\Lambda)$, we have, for $P>0, \ \Lambda\in\mathbb{R}^n$,
\begin{align}\label{Hrela}
H_{2,t}(P,\Lambda) &   =\inf_{\pi\in\mathbb{R}^{m}}\big[P\pi^{\prime}\sigma_{t}\sigma_{t}^{\prime
}\pi-2[P((\pi^{+})^{\prime}\underline{\mu}_{t}-(\pi^{-})^{\prime}\bar{\mu}%
_{t})+\pi^{\prime}\sigma_{t}\Lambda]\big]\nonumber\\
&  =\inf_{\pi\in\mathbb{R}^{m}}\sup_{ v\in\mathcal{B}}\big[P\pi^{\prime}\sigma_{t}\sigma_{t}^{\prime}\pi-2\pi^{\prime
}(Pv+\sigma\Lambda)\big]\nonumber\\
&  =\sup_{v\in\mathcal{B}}\inf_{\pi\in
\mathbb{R}^{m}}\big[P\pi^{\prime}\sigma_{t}\sigma_{t}^{\prime}\pi-2\pi
^{\prime}(Pv+\sigma\Lambda)\big]\nonumber\\
&  =\sup_{v\in\mathcal{B}}\big[-P(v+\frac{\sigma_t\Lambda}{P})'(\sigma_t\sigma_t')^{-1}(v+\frac{\sigma_t\Lambda}{P})\big]\nonumber\\
&  =-\inf_{v\in\mathcal{B}}\big[P(v+\frac{\sigma_t\Lambda}{P})'(\sigma_t\sigma_t')^{-1}(v+\frac{\sigma_t\Lambda}{P})\big],
\end{align}
where we use the min-max theorem in the third equity.

Consider the BSDE with quadratic growth
\begin{equation}
\tilde{Y}_{t}=\int_{t}^{T}g_s(\tilde{Z}_{s})ds-\int_{t}^{T}\tilde{Z}^{\prime
}_{s}dW_{s}, \label{BSDEqua2}%
\end{equation}
where%
\begin{equation}\label{def-g}
g_t(Z):=\inf_{v\in\mathcal{B}} \big|\sigma_t'(\sigma_t\sigma'_t)^{-1}v- Z\big|^2- Z'(I_n-\sigma'_t(\sigma_t\sigma'_t)^{-1}\sigma_t) Z-\frac{1}{2}| Z|^2-2r_t.
\end{equation}
By Theorem 9.6.3 in \cite{CZ}, BSDE (\ref{BSDEqua2}) has a unique solution
$(\tilde{Y},\tilde{Z})\in L^{\infty}(0,T;\mathbb{R})\times\BMO$.

Set $(P_t,\Lambda_t)=(e^{-\tilde Y_t}, -\tilde Z_te^{-\tilde Y_t})$, then $P_T=e^{-\tilde Y_T}=1$. And from the boundedness of $\tilde Y_t$, we know $\Lambda\in\BMO$. Applying It$\hat{\mathrm{o}}$'s formula to
$e^{-\tilde{Y}_{t}}$,
\begin{align*}
dP_t&=de^{-\tilde{Y}_{t}}\\
&=-e^{-\tilde{Y}_{t}}\Big[-\inf_{v\in\mathcal{B}} \Big|\sigma_t'(\sigma_t\sigma'_t)^{-1}v- \tilde Z_t\Big|^2+ \tilde Z_t'(I_n-\sigma'_t(\sigma_t\sigma'_t)^{-1}\sigma_t) \tilde Z_t+2r_t\Big]dt-e^{-\tilde{Y}_{t}}\tilde Z_t'dW_t\\
&=-\Big[2r_tP_t-P_t\inf_{v\in\mathcal{B}} \Big|\sigma_t'(\sigma_t\sigma'_t)^{-1}v+\frac{\Lambda_t}{P_t}\Big|^2+\frac{1}{P_t} \Lambda_t'(I_n-\sigma'_t(\sigma_t\sigma'_t)^{-1}\sigma_t) \Lambda_t\Big]dt+\Lambda_t'dW_t\\
&=-\Big[2r_tP_t-P_t\inf_{v\in\mathcal{B}} \Big|\sigma_t'(\sigma_t\sigma'_t)^{-1}(v+\sigma_t\frac{\Lambda_t}{P_t})+(I_n-\sigma_t'(\sigma_t\sigma_t')^{-1}\sigma_t)\frac{\Lambda_t}{P_t}\Big|^2\\
&\qquad\qquad+\frac{1}{P_t} \Lambda_t'(I_n-\sigma'_t(\sigma_t\sigma'_t)^{-1}\sigma_t) \Lambda_t\Big]dt+\Lambda_t'dW_t\\
&=-\Big[2r_tP_t-P_t\inf_{v\in\mathcal{B}} \Big|\sigma_t'(\sigma_t\sigma'_t)^{-1}(v+\sigma_t\frac{\Lambda_t}{P_t})\Big|^2-P_t\big|(I_n-\sigma_t'(\sigma_t\sigma_t')^{-1}\sigma_t)\frac{\Lambda_t}{P_t}\big|^2\\
&\qquad+\frac{1}{P_t} \Lambda_t'(I_n-\sigma'_t(\sigma_t\sigma'_t)^{-1}\sigma_t) \Lambda_t\Big]dt+\Lambda_t'dW_t\\
&=-\Big[2r_tP_t-P_t\inf_{v\in\mathcal{B}} \Big|\sigma_t'(\sigma_t\sigma'_t)^{-1}(v+\sigma_t\frac{\Lambda_t}{P_t})\Big|^2\Big]dt+\Lambda_t'dW_t\\
&=-\Big[2r_tP_t+H_{2,t}(P_t,\Lambda_t)\Big]dt+\Lambda_t'dW_t,
\end{align*}
where we have used    the orthogonality of  $\sigma_t'(\sigma_t\sigma'_t)^{-1}(v+\sigma_t\frac{\Lambda_t}{P_t})$ and $(I_n-\sigma_t'(\sigma_t\sigma_t')^{-1}\sigma_t)\frac{\Lambda_t}{P_t}$ in  the fifth equality, the idempotency of  $I_n-\sigma'_t(\sigma_t\sigma'_t)^{-1}\sigma_t$ in the sixth equality   and \eqref{Hrela} in the last equality.

Note that $\tilde Y$ is bounded, thus there exists a constant $C>0$ such that $P_t=e^{-\tilde Y_t}\geq C$. This shows that $(P_t,\Lambda_t)$ is actually a solution to \eqref{Riccati2}.

\bigskip

Let us now prove the uniqueness. Suppose $(P, \Lambda)$ and $(\tilde P, \tilde\Lambda)$ are two solutions of \eqref{Riccati2}, such that $ P\geq C, \ \tilde P\geq C$ for some $C>0$.
Define the processes
\begin{align*}
(U,V)=\left(\ln P,\frac{\Lambda}{P}\right), \ (\tilde U,\tilde V)=\left(\ln \tilde P,\frac{\tilde \Lambda}{\tilde P}\right).
\end{align*}
Then $(U, V), \ (\tilde U, \tilde V)\in L^{\infty
}(0,T;\mathbb{R})\times \BMO$. By It\^{o}'s formula  and similar analysis as in the proof of the existence, it's not hard to show that both $(U, V)$ and $(\tilde U, \tilde V)$ are solutions of \eqref{BSDEqua2}. From the uniqueness of  solution to \eqref{BSDEqua2}, we have $U=\tilde U$. Hence $P=\tilde P$, which gives the uniqueness of solution to \eqref{Riccati2}.  This completes the proof.
\eof

\begin{remark}
If $m=n$, then $I_n-\sigma'_t(\sigma_t\sigma'_t)^{-1}\sigma_t=0$, and \eqref{def-g} becomes
\begin{align*}
g_t(Z):=\inf_{v\in\mathcal{B}} \big|\sigma_t'(\sigma_t\sigma'_t)^{-1}v- Z\big|^2-\frac{1}{2}| Z|^2-2r_t.
\end{align*}
\end{remark}

 The following corollary is useful in determining the Lagrange
multiplier.

\begin{corollary}\label{cobon}
\label{strictlybound}Suppose Assumptions \ref{assu1}, \ref{assu3} and \eqref{feasible} hold. Let $(P_{1,t},\Lambda_{1,t})$ and $(P_{2,t}%
,\Lambda_{2,t})$ be the unique
solutions to $(\ref{Riccati1})$ and $(\ref{Riccati2})$ respectively. Then we
have
\[
P_{1,0}e^{-2\int_{0}^{T}r_{s}ds}\leq1\text{ and}\ P_{2,0}e^{-2\int_{0}%
^{T}r_{s}ds}<1.
\]
\end{corollary}

\pf By Proposition \ref{boundedness}, we have
$P_{1,0}e^{-2\int_{0}^{T}r_{s}ds}\leq1$ and$\ P_{2,0}e^{-2\int_{0}^{T}%
r_{s}ds}\leq1$.

If $P_{2,0}e^{-2\int_{0}^{T}r_{s}ds}=1$, then $H_{2,t}(P_{2,t},\Lambda
_{2,t})\equiv0$ for $t\in\lbrack0,T]$. Then $(P_{2,t},\Lambda
_{2,t})=(e^{2\int_{t}^{T}r_{s}ds},0)$, which leads to
\[
H_{2,t}(P_{2,t},0)=P_{2,t}\inf_{\pi\in\mathbb{R}^{m}}\Big[\pi^{\prime}%
\sigma_{t}\sigma_{t}^{\prime}\pi-2((\pi^{+})^{\prime}\underline{\mu}_{t}%
-(\pi^{-})^{\prime}\bar{\mu}_{t})\Big]=0.
\]
Note that (\ref{feasible}) implies that either one of the following two
statements hold:

(1) there is at least one of $\underline{\mu}^{i},\ i=1,...,m$ strictly
greater than $0$ on a set of $(t,\omega)$ with strictly positive measure;

(2) there is at least one of $\bar{\mu}^{i},\ i=1,...,m$ strictly lesser than
$0$ on a set of $(t,\omega)$ with strictly positive measure.

Without loss of generality, we suppose that $\underline{\mu}^{1}{\mathbf{1}}_{(t,\omega)\in M}>0$. Then for a.e. a.s. $(t,\omega)\in M$,
\begin{align*}
&  \ \ \ \ \inf_{\pi\in\mathbb{R}^{m}}\Big[\pi^{\prime}\sigma_{t}\sigma
_{t}^{\prime}\pi-2((\pi^{+})^{\prime}\underline{\mu}_{t}-(\pi^{-})^{\prime
}\bar{\mu}_{t})\Big]\\
&  \leq\inf_{\pi\in\mathbb{R}_{+}^{m}}\Big[\pi^{\prime}\sigma_{t}\sigma
_{t}^{\prime}\pi-2\pi^{\prime}\underline{\mu}_{t}\Big]\\
&  \leq\inf_{\pi\in\mathbb{R}_{+}^{m}}\Big[C\pi^{\prime}\pi-2\pi^{\prime
}\underline{\mu}_{t}\Big]\\
&  \leq C(\frac{\underline{\mu}_{t}^{1}}{C},0,...,0)(\frac{\underline{\mu}%
_{t}^{1}}{C},0,...,0)^{\prime}-2(\frac{\underline{\mu}_{t}^{1}}{C}%
,0,...,0)(\underline{\mu}_{t}^{1},\underline{\mu}_{t}^{2},...,\underline{\mu
}_{t}^{m})^{\prime}\\
&  =-\frac{1}{C}(\underline{\mu}_{t}^{1})^{2}<0,
\end{align*}
where $C$ is a strictly positive constant. Thus we deduce a contradiction. This
completes the proof.
\eof

For any $P>0, \ \Lambda\in\mathbb{R}^n$, $H_{1,t}^*(\pi, P,\Lambda)$ is not necessarily convex with respective to $\pi$, so it may admits more than one arguments minimum. Let $\widetilde\Pi_t(P,\Lambda)$ be the set of arguments minimum  of $H_{1,t}^*(\pi, P,\Lambda)$, i.e.
\begin{align*}
\widetilde\Pi_t(P,\Lambda)=\{\pi_{1,t}(P,\Lambda)|H_{1,t}^*(\pi_{1,t}(P,\Lambda), P,\Lambda)=\inf_{\pi\in\mathbb{R}^{m}}H_{1,t}^*(\pi, P,\Lambda)\}.
\end{align*}
Notice that $H_{1,t}^*(\pi, P,\Lambda)$ is continuous with respect to $\pi$, by a measurable selection theorem (see e.g. Corollary 18.14 in \cite{AB} or Proposition 2.4 in \cite{Sc}), there exists a predictable process $\pi_{1,t}(P,\Lambda)\in \widetilde\Pi_t(P,\Lambda)$.
While for any $P>0, \ \Lambda\in\mathbb{R}^n$,
$H_{2,t}^*(\pi, P,\Lambda)$ is strictly convex with respective to $\pi$. So by a measurable selection theorem, it admits a unique predictable argument minimum  $\pi_{2,t}(P,\Lambda)$, such that
\begin{align}
\label{pi2}
\pi_{2,t}(P,\Lambda)  &  =\argmin_{\pi\in\mathbb{R}^{m}%
}\big[P\pi^{\prime}\sigma_{t}\sigma_{t}^{\prime}\pi-2[P((\pi^{+})^{\prime
}\underline{\mu}_{t}-(\pi^{-})^{\prime}\bar{\mu}%
_{t})+\pi^{\prime}\sigma_{t}\Lambda]\big]. %
\end{align}

\begin{theorem}
\label{feedback} Suppose  Assumptions \ref{assu1}, \ref{assu3} and \eqref{feasible} hold. Let $(P_{1,t},\Lambda_{1,t})$ and $(P_{2,t},\Lambda
_{2,t})$ be the unique solutions of $(\ref{Riccati1})$ and $(\ref{Riccati2})$
respectively. For any predictable $\pi_{1,t}\in \widetilde\Pi_{t}(P_{1,t},\Lambda_{1,t})$, $\pi_{2,t}$ defined in \eqref{pi2},  the state feedback control
\begin{equation}
\pi_{t}^d=\pi_{1,t}(P_{1,t},\Lambda_{1,t})\big(X_{t}%
-de^{-\int_{t}^{T}r_{s}ds}\big)^{+}+\pi_{2,t}(P_{2,t},\Lambda
_{2,t})\big(X_{t}-de^{-\int_{t}^{T}r_{s}ds}\big)^{-} \label{pioptimal}%
\end{equation}
is optimal for the problem $(\ref{step1})$. Moreover, the optimal value is
\begin{equation}
\inf_{\pi\in\mathcal{A}(x)}\E(X_{T}-d)^{2}=%
\begin{cases}
P_{1,0}(x-de^{-\int_{0}^{T}r_{s}ds})^{2}, & if\ x\geq de^{-\int%
_{0}^{T}r_{s}ds},\\
P_{2,0}(x-de^{-\int_{0}^{T}r_{s}ds})^{2}, & if\ x\leq de^{-\int%
_{0}^{T}r_{s}ds}.
\end{cases}
\label{costopti}%
\end{equation}

\end{theorem}

\pf
For any $\pi\in{\mathcal{A}}(x)$ with the wealth
process $X$, define
\[
Y_{t}=X_{t}-de^{-\int_{t}^{T}r_{s}ds}.
\]
By Tanaka's formula,
\[
dY_{t}^{+}=I_{\{Y_{t}>0\}}(r_{t}Y_{t}+(\pi_{t}^{+})^{\prime}\underline{\mu}_{t}-(\pi_{t}^{-})^{\prime}\bar{\mu}%
_{t})dt+I_{\{Y_{t}>0\}}\pi_{t}^{\prime}\sigma_{t}dW_{t}+\frac{1}{2}dL_{t},
\]
where $L_{t}$ is the local time of $Y_{t}$ at $0$.

Applying It$\hat{\mathrm{o}}$'s formula to $(Y_{t}^{+})^{2}$, we have
\begin{align*}
&  \ \ \ \ d(Y_{t}^{+})^{2}\\
&  =2Y_{t}^{+}\Big\{I_{\{Y_{t}>0\}}(r_{t}Y_{t}+(\pi_{t}^{+})^{\prime}%
\underline{\mu}_{t}-(\pi_{t}^{-})^{\prime}\bar{\mu
}_{t})dt+I_{\{Y_{t}>0\}}\pi_{t}^{\prime}\sigma_{t}dW_{t}+\frac{1}{2}%
dL_{t}\Big\}+I_{\{Y_{t}>0\}}\pi_{t}^{\prime}\sigma_{t}\sigma_{t}^{\prime}%
\pi_{t}dt\\
&  =\Big\{2r_{t}(Y_{t}^{+})^{2}+2Y_{t}^{+}((\pi_{t}^{+})^{\prime}\underline{\mu}_{t}-(\pi_{t}^{-})^{\prime}\bar{\mu}%
_{t})+I_{\{Y_{t}>0\}}\pi_{t}^{\prime}\sigma_{t}\sigma_{t}^{\prime}\pi
_{t}\Big\}dt+2Y_{t}^{+}\pi_{t}^{\prime}\sigma_{t}dW_{t},
\end{align*}
where we have used the fact $\int_{0}^{t}\mid Y_{t}\mid dL_{t}=0$.
Then applying It$\hat{\mathrm{o}}$'s formula to $P_{1,t}(Y_{t}^{+})^{2}$,
\begin{align}
&  \ \ \ \ \ dP_{1,t}(Y_{t}^{+})^{2}\nonumber\\
&  =\Big\{I_{\{Y_{t}>0\}}P_{1,t}\pi_{t}^{\prime}\sigma_{t}\sigma_{t}^{\prime
}\pi_{t}+2(Y_{t}^{+})\big[P_{1,t}((\pi_{t}^{+})^{\prime}%
\underline{\mu}_{t}-(\pi_{t}^{-})^{\prime}\bar{\mu}_{t}%
)+\pi_{t}^{\prime}\sigma_{t}\Lambda_{1,t}\big]-(Y_{t}^{+})^{2}H_{1,t}%
(P_{1,t},\Lambda_{1,t})\Big\}dt\nonumber\\
&  \ \ \ \  +\Big\{2P_{1,t}Y_{t}^{+}\pi_{t}^{\prime}\sigma_{t}+(Y_{t}^{+}%
)^{2}\Lambda_{1,t}^{\prime}\Big\}dW_{t}. \label{P1positive}%
\end{align}
Similarly,
\begin{align}
&  \ \ \ \ \ dP_{2,t}(Y_{t}^{-})^{2}\nonumber\\
&  =\Big\{I_{\{Y_{t}\leq0\}}P_{2,t}\pi_{t}^{\prime}\sigma_{t}\sigma
_{t}^{\prime}\pi_{t}-2(Y_{t}^{-})\big[P_{2,t}((\pi_{t}^{+})^{\prime}%
\underline{\mu}_{t}-(\pi_{t}^{-})^{\prime}\bar{\mu
}_{t})+\pi_{t}^{\prime}\sigma_{t}\Lambda_{2,t}\big]-(Y_{t}^{-})^{2}H_{2,t}%
(P_{2,t},\Lambda_{2,t})\Big\}dt\nonumber\\
&  \ \ \ \  +\Big\{-2P_{2,t}Y_{t}^{-}\pi_{t}^{\prime}\sigma_{t}+(Y_{t}%
^{-})^{2}\Lambda_{2,t}^{\prime}\Big\}dW_{t}. \label{P2negative}%
\end{align}

For $n\geq1$, define a stopping time $\tau_{n}$ as follows:
\begin{equation}
\tau_{n}=\inf\{t>0\big|\int_{0}^{t}|2P_{1,s}Y_{s}^{+}\sigma_{s}^{\prime}%
\pi_{s}+(Y_{s}^{+})^{2}\Lambda_{1,s}|^{2}ds+\int_{0}^{t}|-2P_{2,s}Y_{s}%
^{-}\sigma_{s}^{\prime}\pi_{s}+(Y_{s}^{-})^{2}\Lambda_{2,s}|^{2}ds\geq
n\}\wedge T, \label{tau}%
\end{equation}
where $\inf\varnothing:=+\infty$. It is obvious that $\{\tau_{n}\}_{n\geq1}$ is an
increasing sequence and converges to $T$. Adding and integrating $(\ref{P1positive}%
)$ and $(\ref{P2negative})$ from $0$ to $\tau_{n}$, we get
\begin{align}
&  \ \ \ \ {\mathbb{E}}\big[P_{1,\tau_{n}}(Y_{\tau_{n}}^{+})^{2}+P_{2,\tau_{n}}%
(Y_{\tau_{n}}^{-})^{2}\big]\nonumber\\
&  =P_{1,0}(Y_{0}^{+})^{2}+P_{2,0}(Y_{0}^{-})^{2}\nonumber\\
&  +{\mathbb{E}}\int_{0}^{\tau_{n}}\Big\{I_{\{Y_{t}>0\}}P_{1,t}\pi
_{t}^{\prime}\sigma_{t}\sigma_{t}^{\prime}\pi_{t}+2(Y_{t}^{+})\big[P_{1,t}%
((\pi_{t}^{+})^{\prime}\underline{\mu}_{t}-(\pi_{t}%
^{-})^{\prime}\bar{\mu}_{t})+\pi_{t}^{\prime}\sigma_{t}%
\Lambda_{1,t}\big]-(Y_{t}^{+})^{2}H_{1,t}(P_{1,t},\Lambda_{1,t})\nonumber\\
&  \ \ \ +I_{\{Y_{t}\leq0\}}P_{2,t}\pi_{t}^{\prime}\sigma_{t}\sigma
_{t}^{\prime}\pi_{t}-2(Y_{t}^{-})\big[P_{2,t}((\pi_{t}^{+})^{\prime}%
\underline{\mu}_{t}-(\pi_{t}^{-})^{\prime}\bar{\mu
}_{t})+\pi_{t}^{\prime}\sigma_{t}\Lambda_{2}\big]-(Y_{t}^{-})^{2}H_{2,t}%
(P_{2,t},\Lambda_{2,t})\Big\}dt. \label{sumup}%
\end{align}

For $t\in\lbrack0,T]$, denote by $\phi(Y_{t},\pi_{t})$ the integrand on the
RHS of the above equation (\ref{sumup}). For any $\pi
\in{\mathcal{A}(x)}$ with the wealth process $X$, define  a $\mathbb{R}^{m}$-valued process $u_{t}$ by
\begin{align*}
u_{t}=
\begin{cases}
\frac{\pi_t}{|Y_t|}, & \text{ if }\ \ Y_{t}\neq0;\\
0, &  \ \text{if }\ \ Y_{t}=0.
\end{cases}
\end{align*}
When $Y_{t}>0$, the drift term on the RHS of
$(\ref{P1positive})$ becomes
\begin{align*}
&  \ \ \ \ P_{1,t}\pi_{t}^{\prime}\sigma_{t}\sigma_{t}^{\prime}\pi_{t}%
+2Y_{t}\big[P_{1,t}((\pi_{t}^{+})^{\prime}\underline{\mu}%
_{t}-(\pi_{t}^{-})^{\prime}\bar{\mu}_{t})+\pi_{t}^{\prime}%
\sigma_{t}\Lambda_{1,t}\big]-Y_{t}^{2}H_{1,t}(P_{1,t},\Lambda_{1,t})\\
&  =Y_{t}^{2}\big\{P_{1,t}u_{t}^{\prime}\sigma_{t}\sigma_{t}^{\prime}%
u_{t}+2\big[P_{1,t}((u_{t}^{+})^{\prime}\underline{\mu}%
_{t}-(u_{t}^{-})^{\prime}\bar{\mu}_{t})+\pi_{t}^{\prime}%
\sigma_{t}\Lambda_{1,t}\big]-H_{1,t}(P_{1,t},\Lambda_{1,t})\big\}\\
&  \geq0
\end{align*}
by the definition of $H_{1,t}(P,\Lambda)$. By the definition of $H_{2,t}%
(P,\Lambda)$, we can show $\phi(Y_{t},\pi_{t})\geq0$ if $Y_t<0$.  Thus, we obtain that $\phi(Y_{t},\pi_{t})$ is nonnegative.

For any $\pi\in\mathcal{A}(x)$, it's easy to verify ${\mathbb{E}}\Big[
\sup\limits_{t\in\lbrack0,T]}|Y_{t}|^{2}\Big]  <\infty$. Let $n\rightarrow
\infty$, and by the dominated convergence theorem, we have
\begin{align*}
{\mathbb{E}}(X_{T}-d)^{2}={\mathbb{E}}(Y_{T})^{2}  &  ={\mathbb{E}}%
\big[P_{1,T}(T)(Y_{T}^{+})^{2}+P_{2,T}(Y_{T}^{-})^{2}%
\big]\\
&  =P_{1,0}(Y_{0}^{+})^{2}+P_{2,0}(Y_{0}^{-})^{2}+{\mathbb{E}}\Big[
\int_{0}^{T}\phi(Y_{t},\pi_{t})dt\Big] \nonumber\\
&  \geq P_{1,0}(Y_{0}^{+})^{2}+P_{2,0}(Y_{0}^{-})^{2},
\end{align*}
where the equality holds at
\[
\pi^d_{t}=\pi_{1,t}(P_{1,t},\Lambda_{1,t})\big(X_{t}%
-de^{-\int_{t}^{T}r_{s}ds}\big)^{+}+\pi_{2,t}(P_{2,t},\Lambda
_{2,t})\big(X_{t}-de^{-\int_{t}^{T}r_{s}ds}\big)^{-},
\]
which is $(\ref{pioptimal})$. As a consequence, $(\ref{costopti})$ is proved.

It remains to prove $\sigma^{\prime}\pi^d\in L^{2}(0,T;\mathbb{R}^{n})$.
Note that
\[
(\pi^{d})^{+}=\pi_{1}^{+}Y^{+}+\pi_{2}^{+}Y^{-},\ \text{and}\ (\pi^{d
})^{-}=\pi_{1}^{-}Y^{+}+\pi_{2}^{-}Y^{-}.
\]
We next prove that the following equation (\ref{wealth2}) has a unique
continuous $\mathcal{F}_{t}$-adapted solution.
\begin{equation}%
\begin{cases}
dY_{t} & =(r_{t}Y_{t}+((\pi_{t}^{d})^{+})^{\prime}%
\underline{\mu}_{t}-((\pi_{t}^{d})^{-})^{\prime}\bar{\mu
}_{t})dt+(\pi_{t}^{d})^{\prime}\sigma_{t}dW_{t}\\
& =(r_{t}Y_{t}+Y_{t}^{+}(\pi_{1}^{+})^{\prime}\underline{\mu}%
_{t}-Y_{t}^{+}(\pi_{1}^{-})^{\prime}\bar{\mu}_{t}+Y_{t}^{-}%
(\pi_{2}^{+})^{\prime}\underline{\mu}_{t}-Y_{t}^{-}(\pi_{2}%
^{-})^{\prime}\bar{\mu}_{t})dt+(Y_{t}^{+}\pi_{1}^{\prime}%
\sigma_{t}+Y_{t}^{-}\pi_{2}^{\prime}\sigma_{t})dW_{t},\\
Y_{0} & =x-de^{-\int_{0}^{T}r_{s}ds},\;t\in\lbrack0,T].
\end{cases}
\label{wealth2}%
\end{equation}
Consider the following two equations:
\begin{align}%
\begin{cases}
d\bar{Y}_{t}=(r_{t}\bar{Y}_{t}+(\pi_{1}^{+})^{\prime}%
\underline{\mu}_{t}\bar{Y}_{t}-(\pi_{1}^{-})^{\prime}\bar{\mu
}_{t}\bar{Y}_{t})dt+\bar{Y}_{t}\pi_{1}^{\prime}\sigma_{t}dW_{t},\\
\bar{Y}_{0}=(x-de^{-\int_{0}^{T}r_{s}ds})^{+},\;t\in\lbrack0,T],
\end{cases}
\label{barY}%
\end{align}
and
\begin{align}%
\begin{cases}
d\tilde{Y}_{t}=(r_{t}\tilde{Y}_{t}-(\pi_{2}^{+})^{\prime}%
\underline{\mu}_{t}\tilde{Y}_{t}+(\pi_{2}^{-})^{\prime}%
\bar{\mu}_{t}\tilde{Y}_{t})dt-\tilde{Y}_{t}\pi_{2}^{\prime}\sigma_{t}%
dW_{t},\\
\tilde{Y}_{0}=(x-de^{-\int_{0}^{T}r_{s}ds})^{-},\;t\in\lbrack0,T].
\end{cases}
\label{tildeY}%
\end{align}
Then
\begin{equation}
\bar{Y}_{t}=(x-de^{-\int_{0}^{T}r_{s}ds})^{+}\exp\Big\{\int_{0}%
^{t}\big(r_{s}+(\pi_{1}^{+})^{\prime}\underline{\mu}_{s}-(\pi
_{1}^{-})^{\prime}\bar{\mu}_{s}-\frac{1}{2}\pi_{1}^{\prime}%
\sigma_{s}\sigma_{s}^{\prime}\pi_{1}\big)ds+\int_{0}^{t}\pi_{1}^{\prime}%
\sigma_{s}dW_{s}\Big\}, \label{solution--explicit-1}%
\end{equation}
and
\begin{equation}
\tilde{Y}_{t}=(x-de^{-\int_{0}^{T}r_{s}ds})^{-}\exp\Big\{\int_{0}%
^{t}\big(r_{s}-(\pi_{2}^{+})^{\prime}\underline{\mu}_{s}+(\pi
_{2}^{-})^{\prime}\bar{\mu}_{s}-\frac{1}{2}\pi_{2}^{\prime}%
\sigma_{s}\sigma_{s}^{\prime}\pi_{2}\big)ds+\int_{0}^{t}\pi_{2}^{\prime}%
\sigma_{s}dW_{s}\Big\}. \label{solution--explicit-2}%
\end{equation}
It's easy to verify that $Y=\bar{Y}-\tilde{Y}$ is a solution of (\ref{wealth2}%
). To prove the uniqueness, let $Y$ and $\dot{Y}$ be two solutions of
(\ref{wealth2}). Set
\[
\hat{Y}_{t}=Y_{t}-\dot{Y}_{t},\ a_{t}=\frac{Y_{t}^{+}-\dot{Y}_{t}^{+}}%
{Y_{t}-\dot{Y}_{t}}I_{\{Y_{t}\neq\dot{Y}_{t}\}},\ b_{t}=\frac{Y_{t}^{-}%
-\dot{Y}_{t}^{-}}{Y_{t}-\dot{Y}_{t}}I_{\{Y_{t}\neq\dot{Y}_{t}\}}.
\]
Then $\hat{Y}$ solves the following linear SDE
\begin{align*}%
\begin{cases}
d\hat{Y}_{t}=\hat{Y}_{t}(r_{t}+a_{t}(\pi_{1}^{+})^{\prime}\underline{\mu}_{t}-a_{t}(\pi_{1}^{-})^{\prime}\bar{\mu
}_{t}+b_{t}(\pi_{2}^{+})^{\prime}\underline{\mu}_{t}-b_{t}%
(\pi_{2}^{-})^{\prime}\bar{\mu}_{t})dt+\hat{Y}_{t}(a_{t}\pi
_{1}^{\prime}\sigma_{t}+b_{t}\pi_{2}^{\prime}\sigma_{t})dW_{t},\\
\hat{Y}_{0}=0,\;t\in\lbrack0,T],
\end{cases}
\end{align*}
which has a unique solution $\hat{Y}=0$.

Thus, (\ref{wealth2}) has a unique solution. We denote it by $Y^{d}$. Then
\[
\pi_{t}^{d}=\pi_{1,t}(P_{1,t},\Lambda_{1,t})({Y_{t}^{d}})^++\pi
_{2,t}(P_{2,t},\Lambda_{2,t})({Y_{t}^{d}})^-.
\]

Denote by $\tau_{n}^{d}$ the stopping time defined in (\ref{tau}) for
$(Y_{t}^{d},\pi_{t}^{d})$. It follows from (\ref{sumup}) that
\begin{equation}
{\mathbb{E}}\big[P_{1,\tau_{n}^{d}}(Y_{\tau_{n}^{d}}^{+})^{2}%
+P_{2,\tau_{n}^{d}}(Y_{\tau_{n}^{d}}^{-})^{2}\big]=P_{1,0}(Y_{0}%
^{+})^{2}+P_{2,0}(Y_{0}^{-})^{2}. \label{sumtau}%
\end{equation}
Recall that from Theorem \ref{riccatipr}, there exists a constant
$C>0$ such that
\[
P_{1,t}\geq C,\ P_{2,t}\geq C,\ t\in\lbrack0,T].
\]
Then by (\ref{sumtau}), we know
\[
C{\mathbb{E}}(Y_{\tau_{n}^{d}\wedge\iota}^{d})^{2}\leq
P_{1,0}(Y_{0}^{+})^{2}+P_{2,0}(Y_{0}^{-})^{2}
\]
for any stopping time  $\iota$ valued in $[0,T]$. Fatou's lemma gives ${\mathbb{E}}(Y_{\iota}^{d})^{2}\leq C$.
By  It$\hat{\mathrm{o}}$'s formula, we have
\begin{align*}
(Y_t^d)^2=y^2+\int_0^t(2r_{s}(Y_{s}^d)^2+2Y_s^d(((\pi_{s}^{d})^{+})^{\prime}%
\underline{\mu}_{s}-((\pi_{s}^{d})^{-})^{\prime}\bar{\mu
}_{s})+|\sigma_s'\pi^d_s|^2)ds+\int_0^t2Y_s^d(\pi_{s}^{d})^{\prime}\sigma_{s}dW_{s}.
\end{align*}
By the definitions of $\pi_{1}$ and $\pi_{2}$, for each $(t,\omega)\in
\lbrack0,T]\times\Omega$, $\pi_{1}^{i}$ and $\pi_{2}^{i}$
take values from $\{0,(-(\sigma\sigma^{\prime})^{-1}(\mu^{I_{1}}+\frac
{\sigma\Lambda_{1}}{P_{1}}))^{i},(\sigma\sigma^{\prime})^{-1}(\mu^{I_{2}}%
+\frac{\sigma\Lambda_{2}}{P_{2}})^{i}:I_{1}\subset\{1,2...,m\},I_{2}%
\subset\{1,2...,m\}\}$. Thus, there exists a constant $C$ such that
\begin{equation}
\int_{0}^{T}|\sigma_{t}^{\prime}\pi_{t}^{d}|^{2}dt\leq C\sup_{0\leq t\leq
T}|Y_{t}^{d}|^{2}\sum_{\substack{I_{1}\subset\{1,2...,m\}\\I_{2}%
\subset\{1,2...,m\}}}\int_{0}^{T}(|(\sigma\sigma^{\prime})^{-1}(\mu^{I_{1}}%
+\frac{\sigma\Lambda_{1}}{P_{1}})|^{2}+|(\sigma\sigma^{\prime})^{-1}(\mu^{I_{2}%
}+\frac{\sigma\Lambda_{2}}{P_{2}})|^{2})dt<+\infty,\ a.s. \label{sigmapi}%
\end{equation}
since $\Lambda_{1}$, $\Lambda_{2}$ are square integrable and the other terms
are bounded. For $n\geq1$, define a stopping time%
\[
\delta_{n}=\inf\{t>0\big|\int_{0}^{t}|Y_s^d\sigma_{s}'\pi_{s}^{d}|^{2}ds\geq
n\}\wedge T.
\]
Then it converges to $T$ almost surely due to (\ref{sigmapi}).
So
\begin{align*}
y^2+\E\int_0^{\tau_n^d\wedge\delta_n}|\sigma_s'\pi^d_s|^2ds=\E(Y_{\tau_n^d\wedge\delta_n}^d)^2
-\E\int_0^{\tau_n^d\wedge\delta_n}(2r_{s}(Y_{s}^d)^2+2Y_s^d(((\pi_{s}^{d})^{+})^{\prime}%
\underline{\mu}_{s}-((\pi_{s}^{d})^{-})^{\prime}\bar{\mu
}_{s}))ds.
\end{align*}
Let $\varepsilon>0$ be the constant in Assumption \ref{assu3}, then we have
\begin{align*}
\varepsilon\E\int_0^{\tau_n^d\wedge\delta_n}|\pi^d_s|^2ds&\leq C+C\E\int_0^{\tau_n^d\wedge\delta_n}2|Y_s^d||\pi_s^d|ds\\
&\leq C+\frac{\varepsilon}{2}\E\int_0^{\tau_n^d\wedge\delta_n}|\pi_s^d|^2ds
+\frac{2C^2}{\varepsilon}\E\int_0^{\tau_n^d\wedge\delta_n}|Y_s^d|^2ds.
\end{align*}
After rearrangement, it follows from Fatou's lemma that
\[
\E\int_0^T|\pi_s^d|^2ds\leq C.
\]
This completes the proof.
\eof
\begin{remark}\label{vertex}
From \eqref{solution--explicit-1} and \eqref{solution--explicit-2}, we can see that if initial wealth $x\leq d^{-\int_0^T r_sds}$, the optimal state process of problem \eqref{step1} will never exceed $d^{-\int_t^T r_sds}$. The case $x\geq d^{-\int_0^T r_sds}$ is parallel.
\end{remark}

\section{Solution to the problem (\ref{maxmin})}
As $d=\lambda+K$, with a slight abuse of notation, both $\lambda$ and $d$ are called Lagrange multipliers in the following.
From  \eqref{lambdatod} and the definition of $\ell$ in \eqref{def:g},
\begin{align*}
\sup_{\lambda\in\R}\ell(\lambda)&=\sup_{\lambda\in\R}\inf_{\pi\in\mathcal{A}(x)}\Big[\E(X_T^{\pi}-K)^2-2\lambda(\E X_T^{\pi}-K)\Big] \\ &=\sup_{d\in\R}\inf_{\pi\in\mathcal{A}(x)}\Big[\E(X_T^{\pi}-d)^2-(d-K)^2\Big].
\end{align*}
Therefore, it is suffices to determine a argument maximum $\hat d\in\R$ of
\[
\sup_{d\in\R}\inf_{\pi\in\mathcal{A}(x)}\Big[\E(X_T^{\pi}-d)^2-(d-K)^2\Big].
\]

From Theorem \ref{feedback},
\begin{align*}
&  \ \ \ \ \inf_{\pi\in\mathcal{A}(x)}\E(X_{T}-d)^{2}-(d-K)^{2}\\
&  =%
\begin{cases}
P_{1,0}(x-de^{-\int_{0}^{T}r_{s}ds})^{2}-(d-K)^{2},\ \mathrm{if}%
\ x\geq de^{-\int_{0}^{T}r_{s}ds};\\
P_{2,0}(x-de^{-\int_{0}^{T}r_{s}ds})^{2}-(d-K)^{2},\ \mathrm{if}%
\ x\leq de^{-\int_{0}^{T}r_{s}ds}.
\end{cases}
\\
&  =%
\begin{cases}
\big(P_{1,0}e^{-2\int_{0}^{T}r_{s}ds}-1\big)d^{2}-\big(2xP_{1,0}%
e^{-\int_{0}^{T}r_{s}ds}-2K\big)d+P_{1,0}x^{2}-K^{2},\ \mathrm{if}%
\ d\leq xe^{\int_{0}^{T}r_{s}ds};\\
\big(P_{2,0}e^{-2\int_{0}^{T}r_{s}ds}-1\big)d^{2}-\big(2xP_{2,0}%
e^{-\int_{0}^{T}r_{s}ds}-2K\big)d+P_{2,0}x^{2}-K^{2},\ \mathrm{if}%
\ d\geq xe^{\int_{0}^{T}r_{s}ds}.
\end{cases}
\end{align*}
Define
\begin{align*}
f(d)  &  =\big(P_{1,0}e^{-2\int_{0}^{T}r_{s}ds}-1\big)d^{2}-\big(2x%
P_{1,0}e^{-\int_{0}^{T}r_{s}ds}-2K\big)d+P_{1,0}x^{2}-K^{2};\\
h(d)  &  =\big(P_{2,0}e^{-2\int_{0}^{T}r_{s}ds}-1\big)d^{2}-\big(2x%
P_{2,0}e^{-\int_{0}^{T}r_{s}ds}-2K\big)d+P_{2,0}x^{2}-K^{2}.
\end{align*}
According to Corollary \ref{strictlybound}, $P_{1,0}e^{-2\int_{0}^{T}r_{s}%
ds}-1\leq0$, $P_{2,0}e^{-2\int_{0}^{T}r_{s}ds}-1<0$. Then we obtain
\begin{align}\label{hmax}
&f(xe^{\int_{0}^{T}%
r_{s}ds})=\max_{d\leq xe^{\int_{0}^{T}r_{s}ds}}f(d)    =-(xe^{\int_{0}^{T}r_{s}ds}-K)^{2}\leq 0,\nonumber\\
&h(\hat d)=\max_{d\geq xe^{\int_{0}^{T}r_{s}ds}}h(d)  = \frac{P_{2,0}e^{-2\int_{0}^{T}r_{s}ds}}{1-P_{2,0}%
e^{-2\int_{0}^{T}r_{s}ds}}\Big(K-xe^{\int_{0}^{T}%
r_{s}ds}\Big)^{2} \geq 0,
\end{align}
where
\begin{equation}
\hat d=\frac{xP_{2,0}e^{-\int_{0}^{T}r_{s}ds}-K}{P_{2,0}e^{-2\int%
_{0}^{T}r_{s}ds}-1}. \label{hatd}%
\end{equation}
Since $K\geq xe^{\int_{0}^{T}r_{s}ds}$, we have
\begin{align}\label{hatdgeq}
\hat d\geq xe^{\int_{0}^{T}r_{s}ds}
\end{align}
and
\[
h(\hat d)\geq 0\geq h(xe^{\int_{0}^{T}r_{s}ds})=-(xe^{\int_{0}^{T}%
r_{s}ds}-K)^{2}.
\]
Thus $\hat d$ defined in \eqref{hatd} is a argument maximum of
\[
\sup_{d\in\R}\inf_{\pi\in\mathcal{A}(x)}\Big[\E(X_T^{\pi}-d)^2-(d-K)^2\Big].
\]


\section{Verification}
For $\pi^d$ and $\hat d$ defined in \eqref{pioptimal} and \eqref{hatd} respectively,
set $\pi^*=\pi^{\hat d}$, then $\pi^*\in\mathcal{A}(x)$ by Theorem \eqref{feedback}, and
\begin{align}\label{lowerbound}
\E(X_T^{\pi^*}-K)^2-2\hat\lambda(\E X_T^{\pi^*}-K)&=\sup_{\lambda\in\R}\Big[\E(X_T^{\pi^\lambda}-K)^2-2\lambda(\E X_T^{\pi^\lambda}-K)\Big]\nonumber\\
&=\sup_{\lambda\in\R}\inf_{\pi\in\mathcal{A}(x)}\Big[\E(X_T^{\pi}-K)^2-2\lambda(\E X_T^{\pi}-K)\Big].
\end{align}
is a lower bound of our original problem \eqref{optm}, noting \eqref{weakdual}.
If we can show $\E X_T^{\pi^*}=K$, then $\pi^*\in\Pi$,  and $\E(X_T^{\pi^*}-K)^2=\E(X_T^{\pi^*}-K)^2-2\hat\lambda(\E X_T^{\pi^*}-K)$ attains the lower bound \eqref{lowerbound} (the LHS of \eqref{weakdual}) which verifies the optimality of $\pi^*$ for problem \eqref{optm}. Thus, it remains to prove $\E X_T^{\pi^*}=K$.
Put $\pi^*=\pi^{\hat d}$ into the wealth equation \eqref{wealth}, and notice that \eqref{wealth2}, \eqref{solution--explicit-2} and
\eqref{hatdgeq}, we have
\begin{align}
\label{X*}
X_t^{\pi^*}&=(x-\hat de^{-\int_{0}^{T}r_{s}ds})\exp\Big\{\int_{0}%
^{t}\big(r_{s}-(\pi_{2}^{+})^{\prime}\underline{\mu}_{s}+(\pi
_{2}^{-})^{\prime}\bar{\mu}_{s}-\frac{1}{2}\pi_{2}^{\prime}%
\sigma_{s}\sigma_{s}^{\prime}\pi_{2}\big)ds\nonumber\\
&\qquad\qquad+\int_{0}^{t}\pi_{2}^{\prime}%
\sigma_{s}dW_{s}\Big\}+\hat de^{-\int_{0}^{t}r_{s}ds},
\end{align}
where $\pi_2$ is given in \eqref{pi2}. As we do not have a explicit expression of $\pi_2$, so it is difficult to verify $\E X_T^{\pi^*}=K$ with the expression \eqref{X*}.

Therefore a more direct expression of the  terminal wealth level under $\pi^*$ is appealing. Noting the convex duality method developed in \cite{CK} for utility maximization problem is efficient in finding the optimal terminal wealth directly. In the following, with $\hat d$ given in \eqref{hatd} and \eqref{hatdgeq}, we will solve the problem \eqref{step1} through convex duality method. As some by products in this procedure, we obtain the variance-optimal martingale measure, a concept firstly introduced in \cite{Sc},  from which the links between the non-linear financial market and classical linear market are obtained. And we find the sub-derivative of the drift in the wealth equation \eqref{wealth} with respect to $\pi$  claimed in Corollary 4.4 of Ji \cite{Ji}.


For any $v\in\mathcal{B}$ (see \eqref{mathcalB} for the definition of $\mathcal{B}$), $\theta\in\BMO$, let  $N_t^{v,\theta}$ be the solution of the following stochastic differential equation,
\begin{align*}
\begin{cases}
dN_{t}^{v,\theta}=-N_{t}^{v,\theta}\Big[r_{t}dt+
\big(\sigma'_t(\sigma_t\sigma'_t)^{-1}v_t+(I_n-\sigma'_t(\sigma_t\sigma'_t)^{-1}\sigma_t)\theta_t\big)'dW_t\Big],\\
N_{0}^{v,\theta}=1.
\end{cases}
\end{align*}
Then $N_t^{v,\theta}e^{\int_0^tr_sds}$ is a uniformly integrable martingale on $[0,T]$. Moreover, the equivalent martingale measures $\{\mathbb{Q}^{v,\theta}\}_{(v,\theta)\in\mathcal{B}\times\BMO}$ in this incomplete market could be constructed by $N_t^{v,\theta}$, i.e.
\[
\frac{d\mathbb{Q}^{v,\theta}}{d\mathbb{P}}\Big|_{\mathcal{F}_T}=N_T^{ v,\theta}e^{\int_0^Tr_sds}.
\]
Note that stochastic exponentials of BMO martingales has been applied to characterize the equivalent martingale measures in Delbaen et al. \cite{DMS}, Choulli et al. \cite{CKS}.

Applying It\^{o}'s formula to $X_sN_{s}^{v,\theta}$ on $[0,t]$, we have
\begin{align}
\label{supermartingale}
X_tN_{t}^{v,\theta}&=x+\int_0^tN_{s}^{v,\theta}\Big[(\pi^+_s)'\underline\mu_s-(\pi^-_s)'\bar\mu_s-\pi_s'v_s\Big]ds\nonumber\\
&\qquad+\int_0^tN_{s}^{v,\theta}\Big[\pi_s'\sigma_s-X_sv_s'(\sigma_s\sigma'_s)^{-1}\sigma_s
-X_s\theta_s'(I_n-\sigma'_s(\sigma_s\sigma'_s)^{-1}\sigma_s)\Big]dW_s.
\end{align}
Set
\begin{align*}
\mathcal{B}_1=\{(v,\theta)\in\mathcal{B}\times\BMO\mid & \  \mbox{the stochastic integral in} \ \eqref{supermartingale} \ \mbox{is a martingale for any} \ \pi\in\mathcal{A}(x)\}.
\end{align*}
Taking expectation of \eqref{supermartingale} and notice that $\underline\mu_t\leq v_t\leq\bar\mu_t$, we have
\begin{align*}
\E[X_TN_{T}^{v,\theta}]\leq x, \ \mbox{for any } \ (v,\theta)\in\mathcal{B}_1, \ \pi\in\mathcal{A}(x).
\end{align*}

\begin{theorem}
\label{convex}
Suppose Assumptions \ref{assu1} and \ref{assu3} hold.
Let $(\tilde{Y},\tilde{Z})$ be the unique solution of (\ref{BSDEqua2}), $\hat d$ defined in \eqref{hatd} and set
\begin{align}\label{hatzeta}
\hat{\zeta}=-2e^{-{\tilde{Y}}_{0}}(x-\hat de^{-\int_{0}^{T}r_{s}ds})
\end{align}
Then the variance-optimal martingale measure $\mathbb{Q}$ is  defined through
$\frac{d\mathbb{Q}}{d\mathbb{P}}\big|_{\mathcal{F}_T}=N_T^{\hat v,\hat\theta}e^{\int_0^Tr_sds}$, where
\begin{equation}
\hat{v}_{t}=\argmin_{ v\in\mathcal{B}}%
|\sigma'_t(\sigma_t\sigma'_t)^{-1}v-\tilde Z_{t}|^{2}, \ \hat\theta_t=\tilde Z_t, \ t\in\lbrack
0,T],\ a.s.\label{sub-derivative}%
\end{equation}
Moreover,  the optimal portfolio of the problem \eqref{step1} could be represented as
\begin{align}
\label{pidual}
\hat\pi_t=-\frac{\hat\zeta}{2}N_{t}^{\hat v,\hat\theta}e^{\tilde Y_t}(\sigma_t\sigma_t')^{-1}(\sigma_t\tilde Z_t-\hat v_t),
\end{align}
and optimal terminal wealth of the problem (\ref{step1}) has the following expression
\[
\hat{X}_T=\hat d-\frac{\hat{\zeta}}{2}N_{T}^{\hat{v},\hat\theta}.
\]
\end{theorem}
\pf
Step 1: Convex duality. Note that
$\hat d\geq xe^{\int_{0}^{T}r_{s}ds}$ in \eqref{hatdgeq} and  Remark \ref{vertex},
the terminal wealth $X^{\pi^{\hat d}}_T$ will never exceed $\hat d$.
For $0<\zeta<\infty$, define
\[
u(\zeta)=\inf\limits_{x\leq \hat d}[(x-\hat d)^{2}+\zeta
x]=\hat d\zeta-\frac{\zeta^{2}}{4}\text{.}%
\]
Then $\forall \pi\in\mathcal{A}(x), \ \forall\,\zeta>0, \ \forall (v,\theta)\in\mathcal{B}_1$, we have
\begin{align*}
\E(X_T^\pi-d)^{2}  &  \geq \E[ u(\zeta N_{T}^{v,\theta})-\zeta X_T^\pi N_{T}^{v,\theta}]\\
&= \E[\hat d\zeta N_{T}^{v,\theta}-\frac
{\zeta^{2}}{4}(N_{T}^{v,\theta})^{2}-\zeta X_T^\pi N_{T}^{v,\theta}]\\
&  \geq \hat d\zeta e^{-\int_{0}^{T}r_{s}ds}-\frac{\zeta^{2}}{4}%
\E(N_{T}^{v,\theta})^{2}-x\zeta
\end{align*}
and the equalities hold if and only if there exists
$\hat\pi\in\mathcal{A}(x), \ \hat{\zeta}>0$, and $(\hat{v},\hat\theta)\in\mathcal{B}_1$, such that
\begin{align*}
X_T^{\hat\pi}=\hat X_T:=\hat d-\frac{\hat{\zeta}}{2}N_{T}^{\hat{v},\hat\theta},
\end{align*}
is the  terminal wealth under the portfolio $\hat\pi$,
and
\begin{equation}
\E[\hat X_TN_{T}^{\hat{v},\hat\theta}]=x \label{bug}%
\end{equation}
holds simultaneously.
So we introduce the dual problem
\begin{align}
&  \sup_{\substack{\zeta>0\\ (v,\theta)\in\mathcal{B}_1}}\ \big[\hat d\zeta e^{-\int_{0}^{T}%
r_{s}ds}-\frac{\zeta^{2}}{4}\E(N_{T}^{v,\theta})^{2}-x\zeta
\big]\nonumber\label{dual}\\
=-  &  \inf_{\substack{\zeta>0\\ (v,\theta)\in\mathcal{B}_1}}\Big[-\hat d\zeta e^{-\int_{0}%
^{T}r_{s}ds}+\frac{\zeta^{2}}{4}\E(N_{T}^{v,\theta})^{2}+x\zeta
\Big]\nonumber\\
=-  &  \inf_{\zeta>0}\Big[\frac{\zeta^{2}}{4}\inf_{(v,\theta)\in\mathcal{B}_1}%
\E(N_{T}^{v,\theta})^{2}+\zeta(x-\hat de^{-\int_{0}^{T}r_{s}ds})\Big].
\end{align}

We first deal with the term $\inf\limits_{(v,\theta)\in\mathcal{B}_1}\E(N_{T}^{v,\theta})^{2}$.
From the definitions of $N_{t}^{v,\theta}$ and $\tilde{Y}_{t}$ (see \eqref{BSDEqua2}),
\begin{align*}
(N_{t}^{v,\theta})^{2}e^{\tilde{Y}_{t}}  =e^{\tilde{Y}_{0}}& \exp\Big\{\int_0^t\Big[\tilde Z-2\sigma'(\sigma\sigma')^{-1}v-2(I_n-\sigma'(\sigma\sigma')^{-1}\sigma)\theta\Big]'dW_s\\
&\qquad\qquad-\frac{1}{2}\int_0^t\big|\tilde Z-2\sigma'(\sigma\sigma')^{-1}v
-2(I_n-\sigma'(\sigma\sigma')^{-1}\sigma)\theta\big|^2ds\Big\}\\
&\cdot\exp\Big\{\frac{1}{2}\int_0^t\big|\tilde Z-2\sigma'(\sigma\sigma')^{-1}v
-2(I_n-\sigma'(\sigma\sigma')^{-1}\sigma)\theta\big|^2ds\Big\}\\
&\cdot\exp\Big\{\int_0^t\Big[-\big|\sigma'(\sigma\sigma')^{-1}v
+(I_n-\sigma'(\sigma\sigma')^{-1}\sigma)\theta\big|^2-2r-g(\tilde Z)\Big]ds\Big\}\\
=e^{\tilde{Y}_{0}}& \exp\Big\{\int_0^t\Big[\tilde Z-2\sigma'(\sigma\sigma')^{-1}v-2(I_n-\sigma'(\sigma\sigma')^{-1}\sigma)\theta\Big]'dW_s\\
&\qquad\qquad-\frac{1}{2}\int_0^t\big|\tilde Z-2\sigma'(\sigma\sigma')^{-1}v
-2(I_n-\sigma'(\sigma\sigma')^{-1}\sigma)\theta\big|^2ds\Big\}\\
&\cdot\exp\Big\{\int_0^t\Big[\big|\sigma'(\sigma\sigma')^{-1}v-\tilde Z\big|^2+(\theta-\tilde Z)'(I_n-\sigma'(\sigma\sigma')^{-1}\sigma)(\theta-\tilde Z)\Big] ds\Big\}\\
&\cdot\exp\Big\{\int_0^t\Big[-\tilde Z'(I_n-\sigma'(\sigma\sigma')^{-1}\sigma)\tilde Z-\frac{1}{2}|\tilde Z|^2-2r-g(\tilde Z)\Big]ds\Big\}.
\end{align*}
From the definition of $g$ (\ref{def-g}), $(N_{t}^{v,\theta})^{2}e^{\tilde{Y}_{t}}$ is a submartingale
for any $(v,\theta)\in\mathcal{B}_1$. By the martingale principle  \cite{El}, $(\hat
{v}, \hat\theta)\in\mathcal{B}_1$ is an optimal solution of $\inf\limits_{(v,\theta)\in\mathcal{B}_1}\E(N_{T}^{v,\theta})^{2}$ if and only if $(N_{t}^{\hat{v},\hat\theta})^{2}e^{\tilde
{Y}_{t}}$ is a martingale. Then we get the representation of $(\hat v, \hat\theta)$ in \eqref{sub-derivative}:
\[
\hat{v}_{t}=\argmin_{ v\in\mathcal{B}}%
|\sigma'_t(\sigma_t\sigma'_t)^{-1}v-\tilde Z_{t}|^{2}, \ \hat\theta_t=\tilde Z_t,\ t\in[0,T],\ a.s.
\]
and
\[
\inf\limits_{(v,\theta)\in\mathcal{B}_1}\E(N_{T}^{v,\theta})^{2}=e^{\tilde{Y}_{0}}.
\]
By simple calculation, the first infimum in (\ref{dual}) is attained at
\[
\hat{\zeta}=-2e^{-{\tilde{Y}}_{0}}(x-\hat de^{-\int_{0}^{T}r_{s}ds})>0.
\]
Clearly $\hat X_T=\hat d-\frac{\hat{\zeta}}{2}%
N_{T}^{\hat{v},\hat\theta}$ satisfies (\ref{bug}).

Step 2: We will show
that there exists a portfolio $\hat\pi\in\mathcal{A}(x)$ such that $X^{\hat\pi}_T=\hat X_T$.

Define a $\mathcal{F}_t$-adapted process $\hat X$ via
\begin{align*}
\hat X_t N_{t}^{\hat v,\hat\theta}=\E[\hat X_TN_{T}^{\hat v,\hat\theta}|\mathcal{F}_t] &= \E[(\hat d-\frac{\hat{\zeta}}{2}%
N_{T}^{\hat{v},\hat\theta})N_{T}^{\hat v,\hat\theta}|\mathcal{F}_t]\nonumber\\
&=\hat de^{-\int_0^Tr_sds}e^{\int_0^tr_sds}N_{t}^{\hat v,\hat\theta}
-\frac{\hat\zeta}{2}(N_{t}^{\hat v,\hat\theta})^2e^{\tilde Y_t}, \ t\in[0,T],
\end{align*}
i.e.
\begin{align}
\label{martingale}
\hat X_t =\hat de^{-\int_t^Tr_sds}
-\frac{\hat\zeta}{2}N_{t}^{\hat v,\hat\theta}e^{\tilde Y_t}, \ t\in[0,T].
\end{align}
Clearly we have $\hat X_0=x$. And
\begin{align*}
\E[\hat dN_{T}^{\hat v,\hat\theta}|\mathcal{F}_t]&=\hat de^{-\int_0^Tr_sds}e^{\int_0^tr_sds}N_{t}^{\hat v,\hat\theta}\\
&=\hat de^{-\int_0^Tr_sds}\Big[1-\int_0^te^{\int_0^sr_\alpha d\alpha}N_{s}^{\hat v,\hat\theta}\Big(\sigma_s'(\sigma_s\sigma_s')^{-1}\hat v_s+(I_n-\sigma_s'(\sigma_s\sigma_s')^{-1}\sigma_s)\hat\theta_s\Big)'dW_s\Big],
\end{align*}
\begin{align*}
\E[\frac{\hat{\zeta}}{2}%
(N_{T}^{\hat{v},\hat\theta})^2|\mathcal{F}_t]&=\frac{\hat\zeta}{2}(N_{t}^{\hat v,\hat\theta})^2e^{\tilde Y_t}\\
&=\frac{\hat\zeta}{2}\Big[e^{\tilde Y_0}+\int_0^t(N_{s}^{\hat v,\hat\theta})^2e^{\tilde Y_s}\Big(\tilde Z_s-2\sigma_s'(\sigma_s\sigma_s')^{-1}\hat v_s-2(I_n-\sigma_s'(\sigma_s\sigma_s')^{-1}\sigma_s)\hat\theta_s\Big)'dW_s\Big].
\end{align*}
From \eqref{supermartingale} and the above two equations, it suffices to prove that there exists $\hat \pi\in\mathcal{A}(x)$ such that
\begin{align}
\label{diffusion}
&\qquad N_{s}^{\hat v,\hat \theta}\Big[\hat\pi_s'\sigma_s-\hat X_s\hat v_s'(\sigma_s\sigma'_s)^{-1}\sigma_s
-\hat X_s\hat\theta_s'(I_n-\sigma'_s(\sigma_s\sigma'_s)^{-1}\sigma_s)\Big] \nonumber\\
&=-\hat de^{-\int_0^Tr_sds}e^{\int_0^sr_\alpha d\alpha}N_{s}^{\hat v,\hat\theta}\Big(\sigma_s'(\sigma_s\sigma_s')^{-1}\hat v_s+(I_n-\sigma_s'(\sigma_s\sigma_s')^{-1}\sigma_s)\hat\theta_s\Big)'\nonumber\\
&\qquad -\frac{\hat\zeta}{2}(N_{s}^{\hat v,\hat\theta})^2e^{\tilde Y_s}\Big(\tilde Z_s-2\sigma_s'(\sigma_s\sigma_s')^{-1}\hat v_s-2(I_n-\sigma_s'(\sigma_s\sigma_s')^{-1}\sigma_s)\hat\theta_s\Big)',
\end{align}
and
\begin{align}
\label{picondition}
(\hat\pi^+_s)'\underline\mu_s-(\hat\pi^-_s)'\bar\mu_s-\hat\pi_s'\hat v_s=0
\end{align}
hold simultaneously.
Noting \eqref{martingale} and $\hat\theta=\tilde Z$, we have
\begin{align*}
&\qquad -N_{s}^{\hat v,\hat \theta}
\hat X_s\hat\theta_s'(I_n-\sigma'_s(\sigma_s\sigma'_s)^{-1}\sigma_s) =-\hat de^{-\int_0^Tr_sds}e^{\int_0^sr_\alpha d\alpha}N_{s}^{\hat v,\hat\theta}\hat\theta_s'(I_n-\sigma_s'(\sigma_s\sigma_s')^{-1}\sigma_s)\nonumber\\
&\qquad\qquad\qquad -\frac{\hat\zeta}{2}(N_{s}^{\hat v,\hat\theta})^2e^{\tilde Y_s}\Big( (I_n-\sigma_s'(\sigma_s\sigma_s')^{-1}\sigma_s)\tilde Z_s-2(I_n-\sigma_s'(\sigma_s\sigma_s')^{-1}\sigma_s)\hat\theta_s\Big)'.
\end{align*}
Therefore
\eqref{diffusion} is equivalent to
\begin{align}\label{diffusioneq}
N_{s}^{\hat v,\hat \theta}\Big[\hat\pi_s'\sigma_s-\hat X_s\hat v_s'(\sigma_s\sigma'_s)^{-1}\sigma_s\Big]&=-\hat de^{-\int_0^Tr_sds}e^{\int_0^sr_\alpha d\alpha}N_{s}^{\hat v,\hat\theta}\Big(\sigma_s'(\sigma_s\sigma_s')^{-1}\hat v_s\Big)'\nonumber\\
&\qquad -\frac{\hat\zeta}{2}(N_{s}^{\hat v,\hat\theta})^2e^{\tilde Y_s}\Big(\sigma_s'(\sigma_s\sigma_s')^{-1}\sigma_s\tilde Z_s-2\sigma_s'(\sigma_s\sigma_s')^{-1}\hat v_s\Big)'.
\end{align}
Noting \eqref{martingale},  $\hat\pi$ defined in \eqref{pidual}
satisfies \eqref{diffusioneq}, hence \eqref{diffusion}. Moreover, we claim that $\hat\pi$ satisfies \eqref{picondition}. Actually, note that
\begin{align}
\label{vequi}
\hat{v}_{t}&=\argmin_{ v\in\mathcal{B}}%
|\sigma'_t(\sigma_t\sigma'_t)^{-1}v-\tilde Z_{t}|^{2}\nonumber\\
&=\argmin_{ v\in\mathcal{B}}%
|\sigma'_t(\sigma_t\sigma'_t)^{-1}v-\sigma'_t(\sigma_t\sigma'_t)^{-1}\sigma_t\tilde Z_{t}-(I_n-\sigma'_t(\sigma_t\sigma'_t)^{-1}\sigma_t)\tilde Z_t|^{2}\nonumber\\
&=\argmin_{ v\in\mathcal{B}}%
|\sigma'_t(\sigma_t\sigma'_t)^{-1}v-\sigma'_t(\sigma_t\sigma'_t)^{-1}\sigma_t\tilde Z_{t}|^{2}\nonumber\\
&=\argmin_{ v\in\mathcal{B}}%
|\sigma'_t(\sigma_t\sigma'_t)^{-1}(v-\sigma_t\tilde Z_{t})|^{2}.
\end{align}
For any $u\in\mathcal{B}$ and $\varepsilon\in(0,1]$, we have $\hat v+\varepsilon(u-\hat v)\in\mathcal{B}$ because $\mathcal{B}$ is convex, and
\begin{align*}
\frac{1}{\varepsilon}\Big[|\sigma'_t(\sigma_t\sigma'_t)^{-1}(\hat v-\sigma_t\tilde Z_{t})|^{2}- |\sigma'_t(\sigma_t\sigma'_t)^{-1}(\hat v+\varepsilon(u-\hat v)-\sigma_t\tilde Z_{t})|^{2}\Big]\leq0.
\end{align*}
Sending $\varepsilon\downarrow0$, we get
\begin{align}
\label{optmcon}
(u-\hat v)'(\sigma_t\sigma_t')^{-1}(\hat v-\sigma_t\tilde Z_t)\geq0, \ \forall u\in\mathcal{B}.
\end{align}
Denote the $i^{th}$ component of $(\sigma_t\sigma_t')^{-1}(\hat v-\sigma_t\tilde Z_t)$ by $((\sigma_t\sigma_t')^{-1}(\hat v-\sigma_t\tilde Z_t))^i, \ i=1,...,m$. Then there must be
\begin{align*}
\hat v^i=
\begin{cases}
\underline\mu^i, \ \mbox{if} \ ((\sigma_t\sigma_t')^{-1}(\hat v-\sigma_t\tilde Z_t))^i\geq0,\\
\bar\mu^i, \ \mbox{if} \ ((\sigma_t\sigma_t')^{-1}(\hat v-\sigma_t\tilde Z_t))^i\leq0.
\end{cases}
\end{align*}
Recall the presentation \eqref{pidual}, this implies \eqref{picondition}.

Step 3:
We will show that $(\hat v,\hat\theta)\in\mathcal{B}_1$, i.e. the stochastic integral in \eqref{supermartingale}    is a martingale for any $\pi\in\mathcal{A}(x)$. According to \eqref{supermartingale}, it suffices to prove that $$\int_0^tN_{s}^{\hat v,\hat\theta}\Big[\pi_s'\sigma_s-X_s\hat v_s'(\sigma_s\sigma'_s)^{-1}\sigma_s
-X_s\hat\theta_s'(I_n-\sigma'_s(\sigma_s\sigma'_s)^{-1}\sigma_s)\Big]dW_s$$ is a uniformly integrable martingale for any $\pi\in\mathcal{A}(x)$.
Recall that $e^{\int_0^tr_sds}N_{t}^{\hat{v},\hat\theta}$ and $(N_{t}^{\hat{v},\hat\theta})^{2}e^{\tilde{Y}_{t}}$ are two uniformly integrable martingales and $r$ is bounded, we have
\begin{align*}
\E\Big[\sup_{t\in[0,T]}|N_{t}^{\hat{v},\hat\theta}|^{2}\Big]\leq C\E\Big[\sup_{t\in[0,T]}e^{2\int_0^tr_sds}|N_{t}^{\hat{v},\hat\theta}|^{2}\Big]\leq C e^{2\int_0^Tr_sds}\E\Big[|N_{T}^{\hat{v},\hat\theta}|^{2}\Big]=C e^{2\int_0^Tr_sds}e^{\tilde Y_0}<\infty,
\end{align*}
where the second inequality is due to Doob's inequality. Thus we have
\begin{align*}
\E\Big[\Big(\int_0^T|N_t^{\hat v,\hat\theta}\pi_t'\sigma_t|^2dt\Big)^{\frac{1}{2}}\Big]\leq \frac{1}{2}\E\Big[\sup_{t\in[0,T]}|N_{t}^{\hat{v},\hat\theta}|^{2}+\int_0^T|\pi_t'\sigma_t|^2dt\Big]<\infty,
\end{align*}
and
\begin{align*}
\E\Big[\Big(\int_0^T|N_{t}^{\hat v,\hat\theta}X_t\hat v_t'(\sigma_t\sigma'_t)^{-1}\sigma_t|^2dt\Big)^{\frac{1}{2}}\Big]&\leq C\E\Big[\Big(\int_0^T|N_{t}^{\hat v,\hat\theta}X_t|^2dt\Big)^{\frac{1}{2}}\Big]\\
&\leq C\sqrt{T}\E\Big[\sup_{t\in[0,T]}|N_{t}^{\hat v,\hat\theta}X_t|\Big]\\
&\leq \frac{C\sqrt{T}}{2}\E\Big[\sup_{t\in[0,T]}|N_{t}^{\hat v,\hat\theta}|^2+\sup_{t\in[0,T]}|X_t|^2\Big]<\infty.
\end{align*}

From the definition of $N_{t}^{\hat v,\hat\theta}$, we know
\begin{align*}
\int_0^tN_{s}^{\hat v,\hat\theta}
\hat\theta_s'(I_n-\sigma'_s(\sigma_s\sigma'_s)^{-1}\sigma_s)dW_s=1-N_t^{\hat v,\hat\theta}-\int_0^tr_sN_s^{\hat v,\hat\theta}ds-\int_0^tN_s^{\hat v,\hat\theta}\hat v_s'(\sigma_s\sigma_s')^{-1}\sigma_sdW_s.
\end{align*}
By the BDG inequality, we have
\begin{align*}
&\qquad\E\int_0^T|N_{t}^{\hat v,\hat\theta}
\hat\theta_t'(I_n-\sigma'_t(\sigma_t\sigma'_t)^{-1}\sigma_t)|^2dt\\
&\leq C\E\Big[\sup_{t\in[0,T]}\Big|\int_0^tN_{s}^{\hat v,\hat\theta}
\hat\theta_s'(I_n-\sigma'_s(\sigma_s\sigma'_s)^{-1}\sigma_s)dW_s\Big|^2\Big]\\
&\leq C+C\E\Big[\sup_{t\in[0,T]}|N_t^{\hat v,\hat\theta}|^2+\Big|\int_0^TN_s^{\hat v,\hat\theta}ds\Big|^2+\sup_{t\in[0,T]}\Big|\int_0^tN_s^{\hat v,\hat\theta}\hat v_s'(\sigma_s\sigma_s')^{-1}\sigma_sdW_s|^2\Big]\\
&\leq C+C\E\Big[\sup_{t\in[0,T]}\Big|\int_0^tN_s^{\hat v,\hat\theta}\hat v_s'(\sigma_s\sigma_s')^{-1}\sigma_sdW_s\Big|^2\Big]\\
&\leq C+C\E\Big[\int_0^T\Big|N_s^{\hat v,\hat\theta}\hat v_s'(\sigma_s\sigma_s')^{-1}\sigma_s\Big|^2 ds\Big]<\infty.
\end{align*}
Then
\begin{align*}
&\qquad\E\Big[\Big(\int_0^T|N_{t}^{\hat v,\hat\theta}X_t
\hat\theta_t'(I_n-\sigma'_t(\sigma_t\sigma'_t)^{-1}\sigma_t)|^2dt\Big)^{\frac{1}{2}}\Big]\\
&\leq \E\Big[\Big(\sup_{t\in[0,T]}X_t\Big)\Big(\int_0^T|N_{t}^{\hat v,\hat\theta}
\hat\theta_t'(I_n-\sigma'_t(\sigma_t\sigma'_t)^{-1}\sigma_t)|^2dt\Big)^{\frac{1}{2}}\Big]\\
&\leq \frac{1}{2}\E\Big[\Big(\sup_{t\in[0,T]}X_t\Big)^2+\int_0^T|N_{t}^{\hat v,\hat\theta}
\hat\theta_t'(I_n-\sigma'_t(\sigma_t\sigma'_t)^{-1}\sigma_t)|^2dt\Big]<\infty.
\end{align*}
From the BDG inequality,
$$\int_0^tN_{s}^{\hat v,\hat\theta}\Big[\pi_s'\sigma_s-X_s\hat v_s'(\sigma_s\sigma'_s)^{-1}\sigma_s
-X_s\hat\theta_s'(I_n-\sigma'_s(\sigma_s\sigma'_s)^{-1}\sigma_s)\Big]dW_s$$
is actually a uniformly integrable martingale for any $\pi\in\mathcal{A}(x_0)$.

Step 4: We need to show $\hat\pi\in L^2(0,T;\mathbb{R}^m)$. And this can be guaranteed by similar method as in the proof of theorem \ref{feedback} after noticing that
$N_{s}^{\hat v,\hat\theta}e^{\tilde Y_s}$ satisfying the following equation
\begin{align*}
\begin{cases}
d(-\frac{\hat\zeta}{2}N_{s}^{\hat v,\hat\theta}e^{\tilde Y_s})=\Big[-\frac{\hat\zeta}{2}r_sN_{s}^{\hat v,\hat\theta}e^{\tilde Y_s}+\hat\pi'_s\hat v_s\Big]ds+\hat\pi'_s\sigma_sdW_s,\\
-\frac{\hat\zeta}{2}N_{0}^{\hat v,\hat\theta}e^{\tilde Y_0}=-\frac{\hat\zeta}{2}e^{\tilde Y_0}.
\end{cases}
\end{align*}

Step 5: Combine \eqref{hatd} and \eqref{hatzeta}, and notice that $P_{2,t}=e^{-\tilde Y_t}$, we have
\begin{align*}
\E\hat{X}_T=\E\Big[\hat d-\frac{\hat{\zeta}}{2}N_{T}^{\hat{v},\hat\theta}\Big]=\hat d-\frac{\hat{\zeta}}{2}e^{-\int_0^Tr_sds}=K.
\end{align*}
This completes the proof.
\eof

\begin{remark}
From \eqref{optmcon}, if $\underline\mu^i<\hat v^i<\bar\mu^i$, there must be $((\sigma_t\sigma_t')^{-1}(\hat v-\sigma_t\tilde Z_t))^i=0$, and $\hat\pi_t^i=0$ by \eqref{pidual}, i.e. the investor should not invest in the $i$th stock.
\end{remark}
\begin{remark}
Both $\pi^*=\pi^{\hat d}$ and $\hat\pi$ defined in \eqref{pidual}  are solutions of the problem \eqref{step1} (with $\hat d$). They  are identical, i.e. $\pi^*=\hat\pi$, the reason is left to the interested readers.
\end{remark}

So far, we achieve the three steps in solving our original problem \eqref{optm}. Therefore we have
\begin{theorem}
Suppose  Assumptions \ref{assu1} and \ref{assu3} hold. Let $(P_{2,t}%
,\Lambda_{2,t})$ be the unique solutions to $(\ref{Riccati2})$, $\pi_2, \hat d$ defined in \eqref{pi2}, \eqref{hatd}.
The efficient strategy of the problem (\ref{optm}) can be written as a
function of time $t$ and the wealth $X_{t}$:
\begin{align*}
\pi^{\ast}(t,X)=-\pi_{2,t}(P_{2,t}%
,\Lambda_{2,t})\big(X_{t}-\hat de^{-\int_{t}^{T}r_{s}ds}\big),
\end{align*}
or equivalently expressed by \eqref{pidual}.
Moreover, the efficient frontier is
\begin{align}
\label{efficientfr}
\mathrm{Var}(X_{T})=\frac{P_{2,0}e^{-2\int_{0}^{T}r_{s}ds}}{1-P_{2,0}%
e^{-2\int_{0}^{T}r_{s}ds}}\Big({\mathbb{E}}X_{T}-xe^{\int_{0}^{T}%
r_{s}ds}\Big)^{2}.
\end{align}
\end{theorem}
\pf
The efficient frontier \eqref{efficientfr} comes from \eqref{hmax}.
\eof

\begin{remark}
When  $m=n=1$ and $\sigma_{t}>0$, we have
\begin{align}
H_{2,t}(P,\Lambda)  &  =\inf_{\pi\in\mathbb{R}}\big[P\sigma_{t}^{2}\pi
^{2}-2[P(\pi^{+}\underline{\mu}_{t}-\pi^{-}\bar
{\mu}_{t})+\pi\sigma_{t}\Lambda]\big]\nonumber\\
&  =%
\begin{cases}
-\frac{(P\underline{\mu}_{t}+\sigma_t\Lambda)^{2}}{P\sigma_t^2}, & \mbox{if} \ \  \frac{\sigma_t\Lambda
}{P}\geq-\underline{\mu}_{t},\nonumber\\
0, & \mbox{if} \ \ -\bar{\mu}_{t}\leq\frac{\sigma_t\Lambda}{P}\leq-\underline{\mu
}_{t},\nonumber\\
-\frac{(P\bar{\mu}_{t}+\sigma_t\Lambda)^{2}}{P\sigma_t^2}, & \mbox{if} \ \  \frac{\sigma_t\Lambda}{P}%
\leq-\bar{\mu}_{t}.
\end{cases}
\end{align}
and
\begin{align*}
\pi_{2,t}(P,\Lambda)=%
\begin{cases}
\frac{P\underline{\mu}_{t}+\sigma_t\Lambda}{P\sigma^2_{t}}, &
\text{if} \ \ \frac{\sigma_t\Lambda}{P}\geq-\underline{\mu}_{t},\\
0, & \text{if} \ \  -\bar{\mu}_{t}\leq\frac{\sigma_t\Lambda}{P}%
\leq-\underline{\mu}_{t},\\
\frac{P\bar{\mu}_{t}+\sigma_t\Lambda}{P\sigma^2_{t}}, &
\text{if} \ \  \frac{\sigma_t\Lambda}{P}\leq-\bar{\mu}_{t}.
\end{cases}
\end{align*}
In the linear financial market, $\underline\mu=\bar\mu=\mu$, then $\pi^{\ast}(t,X)=0$ if and only if $\frac{\sigma_t\Lambda_t}{P_t}=-\mu_t$. While in our non-linear market, $\pi^{\ast}(t,X)=0$ if and only if $-\bar\mu_t\leq\frac{\sigma_t\Lambda_t}{P_t}\leq-\underline\mu_t$. That is to say, the no-trading region becomes larger.
\end{remark}

\begin{remark}
If $m=n=1$, and $\underline{\mu}_{t},$ $\bar{\mu}_{t},$
$\sigma_{t}$ are deterministic continuous functions on $[0,T]$, $0\leq
\underline{\mu}_{t}\leq\bar{\mu}_{t}$ and $\sigma_{t}>0$. Then the
unique solutions of $(\ref{Riccati1})$ and $(\ref{Riccati2})$ are given by
\[
(P_{1,t},\Lambda_{1,t})=(e^{\int_{t}^{T}(2r_{s}-\frac{\bar{\mu}_{s}}{\sigma_s}%
)ds},0);\ (P_{2,t},\Lambda_{2,t})=(e^{\int_{t}^{T}(2r_{s}-\frac{\underline{\mu
}_{s}}{\sigma_s})ds},0).
\]
We recover the same results in \cite{JS}.
\end{remark}


\begin{remark}
For any $v\in\mathcal{B}$, the following BSDE \eqref{Pv} admits a unique solution $(P^v_t,\Lambda^v_t)\in L^\infty(0,T;\mathbb{R})\times\BMO$, such that $P^v_t\geq C$ for some positive constant $C$ by Theorem 2.2 of \cite{KT}.
\begin{align}
\label{Pv}
\begin{cases}
dP^v_t=-\Big\{rP^v_t-P^v_t(v+\frac{\sigma_t\Lambda^v}{P^v_t})'(\sigma_t\sigma_t')^{-1}(v+\frac{\sigma_t\Lambda^v_t}{P^v_t})\Big\}dt+(\Lambda^v_t)'dW_t,\\
P^v_T=1.
\end{cases}
\end{align}
Actually, \eqref{Pv} is the Riccati equation associated with mean-variance problem under the linear wealth equation:
\begin{align*}
\begin{cases}
dX_t=(r_tX_t+\pi'_tv_t)dt+\pi'_t\sigma_tdW_t,\\
X_0=x.
\end{cases}
\end{align*}
Notice that the solution $(P_{2,t},\Lambda_{2,t})$ of \eqref{Riccati2} is uniformly positive, by Theorem 9.6.7 in \cite{CZ}, we have $P^v_t\leq P_{2,t}$ for any $v\in\mathcal{B}$, thus $\mathrm{ess}\sup_{v\in\mathcal{B}}P^v_t\leq P_{2,t}, \ t\in[0,T], \ a.s.$ Then for $\hat v\in\mathcal{B}$ defined in \eqref{sub-derivative},  we have
\[
 P_{2,t}=P^{\hat v}_t=\underset{v\in\mathcal{B}}{\mathrm{ess}\sup}P^v_t, \ t\in[0,T], \ a.s.
\]
by the uniqueness of \eqref{Riccati2}. Similarly, we can prove
\[
 P_{1,t}=P^{\tilde v}_t=\underset{v\in\mathcal{B}}{\mathrm{ess}\inf}P^v_t, \ t\in[0,T], \ a.s.,
\]
where
\[
\tilde v=\argmin_{v\in\mathcal{B}}\big[-(v+\frac{\sigma_t\Lambda_{1,t}}{P_{1,t}})'(\sigma_t\sigma_t')^{-1}(v+\frac{\sigma_t\Lambda_{1,t}}{P_{1,t}})\big].
\]
\end{remark}

\begin{remark}
Let $\hat v$ be defined in \eqref{sub-derivative}, then the problem (\ref{optm}) is equivalent to the following problem with a linear
wealth equation:
\begin{align*}
&  \mathrm{\mbox{Minimize}}\ \E(X_{T}-K)^{2},\\
&  s.t.%
\begin{cases}
\E X_T=K,\\
\sigma'\pi\in L^{2}(0,T;\mathbb{R}^n),\\
dX_{t}=(r_{t}X_{t}+\pi_{t}^{\prime}\hat{v}_{t})dt+\pi_{t}^{\prime}\sigma
_{t}dW_{t},\\
X_{0}=x.
\end{cases}
\end{align*}
Actually, $\hat{v}$ is the sub-derivative of $(\pi^{+})^{\prime
}\underline{\mu}-(\pi^{-})^{\prime}\bar{\mu}$ claimed in Corollary 4.4 of Ji \cite{Ji}.
\end{remark}

\section{Concluding remarks}
In this paper, we study mean-variance portfolio selection under non-linear wealth dynamics. Different from the linear wealth case, by introducing a Lagrange multiplier, we only have the weak duality \eqref{weakdual}. Therefore,  solutions of  the LHS of \eqref{weakdual} only provide a lower bound for our original problem. After constructing a candidate portfolio $\pi^{\hat d}$ from the LHS of \eqref{weakdual}, we need to verify that $\pi^{\hat d}\in\Pi$, i.e. $\pi$ is feasible (hence optimal) for our original problem \eqref{optm} (or equivalently the RHS of \eqref{weakdual}). This is achieved by the convex duality method which gives a more direct expression of the corresponding terminal wealth. Note that the quadratic cost function is not monotone, a property which is usually required for establishing convex duality. Fortunately, Remark \eqref{vertex} and Eq. \eqref{hatdgeq} render the corresponding terminal wealth $X^{\pi^{\hat d}}_T$ of the candidate portfolio stay below $\hat d$. That is to say, without the analysis in Sections 4 and 5, the convex duality could not be established in Section 6.
Finally, the optimal portfolio, efficient frontier and the variance-optimal martingale measure  are given in closed forms. And we find the links between the non-linear financial market and classical linear market.

Extensions in other directions can be interesting as well. For instance: (1) How to characterize the optimal portfolio of  problems \eqref{optm} or \eqref{step1} when the interest rate $r$ is a stochastic process? (2) Recently,  the general form of the mean-variance efficient frontier has been recently established in $\mathrm{\breve{C}}$ern$\mathrm{\acute{y}}$, Czichowsky and Kallsen \cite{CCK} with stochastic interest rates and even only risky assets. Can we generalize the results in \cite{CCK} to the present setting with non-linear wealth dynamics? (3) Mean-variance portfolio selection when the diffusion term is also non-linear with respect to $\pi$.

\renewcommand{\refname}


{\large References}

\bigskip

\begin{thebibliography}{99}                                                                                               
\bibitem {AB} Aliprantis C, Border K (2006) Infinite dimensional analysis. New York: Springer.


\bibitem {BJPZ}Bielecki T, Jin H, Pliska S, Zhou X (2005) \emph{Continuous-time
mean-variance portfolio selection with bankruptcy prohibition}.  Math. Finance.  15(2): 213-244.

\bibitem{CCK} Cerny A, Czichowsky C, and Kallsen J (2021) \emph{Numeraire-invariant quadratic hedging and mean-variance portfolio allocation.} arXiv preprint 2110.09416

\bibitem {CKa} Cerny A, Kallsen J (2007) \emph{On the structure of general mean-variance hedging strategies}.  Ann. Probab.,  35(4): 1479-1531.

\bibitem {CKS} Choulli T, Krawczyk L, Stricker C (1998). \emph{martingales and their applications in mathematical finance}. Ann. Probab., 26(2): 853-876.

\bibitem {CC}Cuoco D, Cvitanic J (1998) \emph{Optimal consumption choices for a
`large' investor}.  J. Econom. Dynam. Control. 22(3): 401-436.

\bibitem {CK}Cvitanic J, Karatzas I (1992) \emph{Convex duality in constrained portfolio optimization}. Ann. Appl. Probab., 2(4), 767-818.

\bibitem {CZ}Cvitanic  J,  Zhang J (2012) \emph{Contract theory in continuous-time models}, Springer Science and Business Media.


\bibitem {CS} Czichowsky C, Schweizer M (2013) \emph{Cone-constrained continuous-time Markowitz problems}. Ann. Appl. Probab., 2013, 23(2): 764-810.

\bibitem {DMS}    Delbaen F, Monat P, Schachermayer W, Schweizer, Stricker (1997)  \emph{Weighted norm inequalities and hedging in incomplete markets}. Finance Stoch., 1(3): 181-227.

\bibitem {DM} Dellacherie C, Meyer P (1978), \emph{Probabilityes and Potertial} , North-holland Mathematics studies 29.

\bibitem {El}El Karoui N (1981) \emph{Les aspects probabilistes du controle
stochastique}, Lecture Notes in Mathematics, Springer-Verlag. 73-238.

\bibitem {EPQ}El Karoui N, Peng S, Quenez M (1997) \emph{Backward stochastic
differential equations in finance}.  Math. Finance. 7(1): 1-71.

\bibitem {EPQ1}El Karoui N, Peng S, Quenez M (2001) \emph{A dynamic maximum
principle for the optimization of recursive utilities under constraints}.  Ann. Appl. Probab. 11(3): 664-693.

\bibitem {FLL}Fu C, Lavassani L, Li X (2010) \emph{Dynamic mean-variance
portfolio selection with borrowing constraint}. European J. Oper. Res. 200(1): 312-319.

\bibitem {HK}Harrison J, Kreps D (1979) \emph{Martingales and arbitrage in
multiperiod securities markets}.  J. Econom. Theory. 20(3): 381-408.

\bibitem {HP1}Harrison J, Pliska S (1981) \emph{Martingales and stochastic
integrals in the theory of continuous trading}.  Stochastic Process. Appl. 11(3): 215-260.

\bibitem {HP2}Harrison J, Pliska S (1983) \emph{A stochastic calculus model of
continuous trading: complete markets}.  Stochastic Process. Appl. 15(3): 313-316.

\bibitem {HZ}Hu Y, Zhou X (2005) \emph{Constrained stochastic LQ control with
random coefficients, and application to portfolio selection}. SIAM J. Control Optim.  44(2): 444-466.

\bibitem {Ji}Ji S (2010) \emph{Dual method for continuous-time Markowitz's problems
with nonlinear wealth equations}.  J. Math. Anal. Appl. 366(1): 90-100.

\bibitem {JS}Ji S, Shi X (2017) \emph{Explicit solutions for continuous time
mean-variance portfolio selection with nonlinear wealth equations}. Systems Control Lett. 104: 1-4.



\bibitem {JZ}Ji S, Zhou X (2006) \emph{A maximum principle for stochastic optimal
control with terminal state constraints, and its applications}. Commun. Inf. Syst. 6(4): 321-338.

\bibitem {JYZ}Jin H, Yan J, Zhou X (2005) \emph{Continuous-time mean-risk
portfolio selection}.  Ann. Inst. H. Poincare Probab. Statist. 41(3): 559-580.

\bibitem {JK}Jouini E, Kallal H (1995) \emph{Arbitrage in Securities Markets with
Short-Sales Constraints}.  Math. Finance. 5(3): 197-232.

\bibitem {JK01} Jouini E and Kallal H (2001) \emph{Efficient trading strategies in the presence of market frictions}. Rev. Financ. Stud., 14(2): 343-369.

\bibitem {KS} Karatzas I, Shreve S (1998). \emph{Methods of mathematical finance}. New York: Springer.

\bibitem {Ka} Kazamaki N (2006) \emph{Continuous exponential martingales and BMO}. Springer.



\bibitem {KT}Kohlmann M, Tang S (2002) \emph{Global adapted solution of
one-dimensional backward stochastic Riccati equations, with application to the
mean--variance hedging}.  Stochastic Process. Appl.
97(2): 255-288.


\bibitem {LN}Li D, Ng W (2000) \emph{Optimal dynamic portfolio selection:
multiperiod mean-variance formulation}.  Math. Finance. 10(3): 387-406.

\bibitem {LZL}Li X, Zhou X, Lim A (2002) \emph{Dynamic mean-variance portfolio
selection with no-shorting constraints}. SIAM J. Control Optim.  40(5): 1540-1555.




\bibitem {Ma1}Markowitz H (1952) \emph{Portfolio selection}. The Journal of
Finance. 7: 77-91.

\bibitem {Ma2}Markowitz H (1968) \emph{Portfolio selection: efficient
diversification of investments}. Yale University Press.

\bibitem {RY}Revuz D, Yor M (2013) \emph{Continuous martingales and Brownian
motion}. Springer Science and Business Media.

\bibitem {Sc}Schweizer M (1996) \emph{Approximation pricing and the variance-optimal martingale measure}. Ann. Probab.,  24(1): 206-236.

\bibitem {Sc2010} Schweizer M (2010). \emph{Mean variance hedging}. Encyclopedia of quantitative finance.

\bibitem {ZL}Zhou X, Li D(2000) \emph{Continuous-time mean-variance portfolio
selection: A stochastic LQ framework}. Appl. Math. Optim.
42(1): 19-33.
\end{thebibliography}
\end{document}